\documentclass{emulateapj}

\slugcomment{ApJ, in press}

\shorttitle{Stellar Feedback in 30 Doradus}
\shortauthors{LOPEZ ET AL.}

\newcommand{\ltsima}{$\; \buildrel < \over \sim \;$}
\newcommand{\simlt}{\lower.5ex\hbox{\ltsima}}

\newcommand{\ls}{{_<\atop^{\sim}}}

\newcommand{\gs}{{_>\atop^{\sim}}}

\def\arcmin{\hbox{$^\prime$}}
\def\arcsec{\hbox{$^{\prime\prime}$}}

\begin{document}

\title{What Drives the Expansion of Giant HII Regions?: A Study of Stellar Feedback in 30~Doradus}

\author{Laura A. Lopez\altaffilmark{1}, Mark R. Krumholz\altaffilmark{1}, Alberto D. Bolatto\altaffilmark{2}, J. Xavier Prochaska\altaffilmark{1,3}, Enrico Ramirez-Ruiz\altaffilmark{1}}
\altaffiltext{1}{Department of Astronomy and Astrophysics, University of California Santa Cruz, 159 Interdisciplinary Sciences Building, 1156 High Street, Santa Cruz, CA 95064, USA; lopez@astro.ucsc.edu.}
\altaffiltext{2}{Department of Astronomy, University of Maryland, College Park, MD 20742, USA}
\altaffiltext{3}{UCO/Lick Observatory}

\begin{abstract}

Observations show that star formation is an inefficient and slow process. This result can be attributed to the injection of energy and momentum by stars that prevents free-fall collapse of molecular clouds. The mechanism of this stellar feedback is debated theoretically: possible sources of pressure include the classical warm HII gas, the hot gas generated by shock-heating from stellar winds and supernovae, direct radiation of stars, and the dust-processed radiation field trapped inside the HII shell. In this paper, we measure observationally the pressures associated with each component listed above across the giant HII region 30 Doradus in the Large Magellanic Cloud. We exploit high-resolution, multi-wavelengh images (radio, infrared, optical, and X-ray) to map these pressures as a function of position. We find that radiation pressure dominates within 75 pc of the central star cluster, R136, while the HII gas pressure dominates at larger radii. By contrast, the dust-processed radiation pressure and hot gas pressure are generally weak and not dynamically important, although the hot gas pressure may have played a more significant role at early times. Based on the low X-ray gas pressures, we demonstrate that the hot gas is only partially confined and must be leaking out the HII shell. Additionally, we consider the implications of a dominant radiation pressure on the early dynamics of 30 Doradus. 

\end{abstract}


\section{Introduction}

Molecular clouds contain the coolest and densest gas in the Universe, and thus they are the sites where stars form. The physical properties of these clouds set the initial conditions for protostellar collapse and may define the stellar initial mass function (IMF) [\citealt{motte,ts98,onishi}]. The massive stars formed there eventually end in supernova explosions, injecting mechanical energy and chemically enriching the interstellar medium (ISM). Therefore, molecular clouds shape the entire stellar life cycle, and an understanding of their properties and dynamics is key to probe galactic evolution.

Observational evidence shows that star formation is an inefficient and slow process. Only 5--10\% of available molecular cloud mass is converted into stars over the cloud lifetime\footnote{Cloud lifetime is debated contentiously in the literature. Observational estimates range from a single free-fall time \citep{elm00,hartmann,ball07} to several free-fall times \citep{tan06}. However, there is a consensus that only a few percent of gas is converted in either timescale.} \citep{wk97} and only $\sim$2\% of the gas is converted to stars in one free-fall time across several orders of magnitude in density \citep{zuckerman,kt07}. This inefficiency can be attributed to the internal processes of HII regions that disrupt their host molecular clouds (e.g., \citealt{m02,mrk06}), but the mode of this stellar feedback remains uncertain. 

Broadly, there are several possible sources of internal energy and momentum that may drive the dynamics of HII regions: the direct radiation from stars (e.g., \citealt{jijina,km09}), the dust-processed infrared radiation trapped inside an HII shell \citep{thompson,murray,andrews}, the warm gas ionized by massive stars (e.g., \citealt{whit,dale}), the hot gas shock heated by stellar winds and supernovae (e.g., \citealt{yorke,hc09}), and protostellar outflows/jets (e.g., \citealt{quill,cunningham,li,naka,wang}). Each of these mechanisms has been considered individually in the literature, but no observational analyses have ever compared the relative contribution of all these components within HII regions. 

In this paper, we investigate the role of the stellar feedback mechanisms listed above in the giant HII region 30 Doradus in the nearby Large Magellanic Cloud (LMC). Several properties of the LMC make it a favorable target: the LMC's proximity ($\sim$50 kpc) ensures individual point sources can be resolved while maintaining the capability of mapping the diffuse emission at sub-pc scales. Additionally, the LMC has a face-on orientation and a low column density (a few $\times$10$^{21}$ cm$^{-2}$) that limits line-of-sight confusion. Given these advantages, the LMC (and thus 30 Doradus) has been surveyed at many wavelengths at high spatial resolution, and we can exploit these data to compare observationally all the feedback mechanisms and how they vary with position across 30 Doradus. 

The text is structured as follows: $\S$1.1 gives relevant background on the source, 30 Doradus, and describes why this source is a good ``test case'' for our analyses. In $\S$2, we present the multiwavelength data utilized in our work to assess the dynamical role of all the possible stellar feedback mechanisms. $\S$3 outlines how we utilize these images to calculate the pressures associated with each feedback component across 30 Doradus. $\S$4 gives the results from our analyses, and $\S$5 discusses the implications of our findings, including evidence of X-ray gas leakage from the HII region ($\S$5.1) and the role of radiation pressure in HII region dynamics ($\S$5.3). Additionally, we articulate the different ways one can define radiation pressure, and how these definitions can lead to divergent results in $\S$5.2. Finally, we summarize and conclude our analysis in $\S$6.

\subsection{Background on 30 Doradus}

30 Doradus is the most massive and largest HII region in the Local Group. The primary star cluster powering 30 Doradus is NGC 2070, with 2400 OB stars \citep{parker1}, an ionizing photon luminosity $S = 4.5 \times 10^{51}$ photons s$^{-1}$ \citep{wal91}, and a bolometric luminosity of 7.8$\times10^{7} L_{\sun}$ \citep{mal94}. The IMF of NGC 2070 has masses up to 120 M$_{\sun}$ \citep{mh98}, and the stellar population may be the result of several epochs of star formation \citep{wb97}. At the core of NGC 2070 is R136, the densest concentration of very massive stars known, with a central density of 5.5 $\times 10^{4} M_{\sun}$ pc$^{-3}$ \citep{hunter95}; R136 hosts at least 39 O3-type stars and at least 10 WR stars in its core ($\sim$2.5 pc in diameter; \citealt{mh98}).

\begin{figure}
\includegraphics[width=\columnwidth]{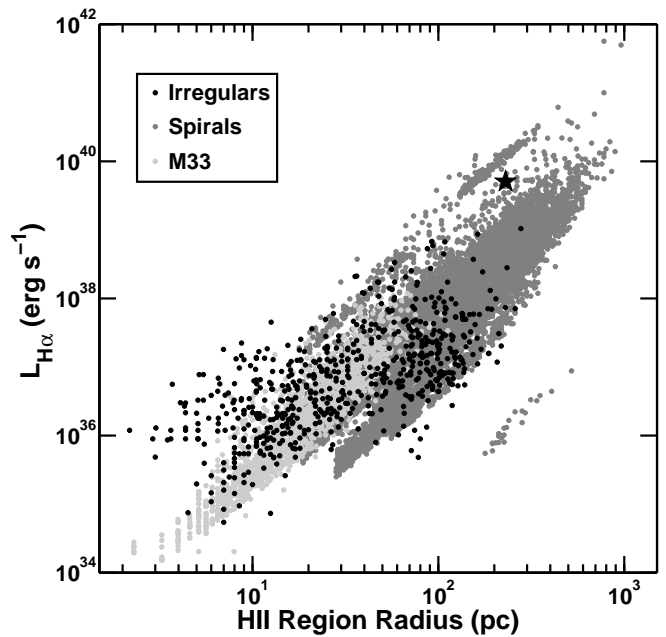}
\caption{H$\alpha$ luminosity versus HII region radius for $\sim$22,000 HII regions in 70 nearby (distances $\ls$30 Mpc) galaxies. The black star at the top right denotes 30 Doradus. It is the brightest HII region in the irregular galaxies by nearly an order of magnitude, and it is more luminous than $\approx$99\% of the HII regions in the spiral galaxies. We note that none of the data is corrected for reddening. The plotted data were compiled from the following references: LMC and SMC: \cite{kenn86}; Sextans~A: \cite{hks94}; NGC~6822: \cite{hlk89}; Holmberg~II: \cite{hsk94}; GR8: \cite{hlk2}; DDO~47, Leo~A, Sextans~B, DDO~167, DDO~168, DDO~187: \cite{shk91}; DDO~53 \cite{shk90}; 56 spirals: \cite{knap}, \cite{bradley}; M33: \cite{wyder}, \cite{h99}. We utilized the distances from \cite{k08} to convert from $H\alpha$ flux to luminosity.}
\label{fig:Ha}
\end{figure}

To provide context for how 30 Doradus compares to other local HII regions, Figure~\ref{fig:Ha} plots H$\alpha$ luminosity versus HII region radius for $\sim$22,000 HII regions in 70 nearby (distances $\ls$30 Mpc) galaxies (see references in figure caption). Morphologically, this galaxy sample is comprised of 13 irregulars/dwarf irregulars and 57 spirals. The black star near the top right denotes 30 Doradus. It is the brighest in H$\alpha$ of the 613 HII regions in the irregulars by nearly an order of magnitude. Relative to the HII regions in spirals (including M33), 30 Doradus has a greater H$\alpha$ luminosity than $\sim$99\% of that sample.  

The nebula that surrounds the central star cluster has a complex morphology across the electromagnetic spectrum. Figure~\ref{fig:threecolor} shows a three-color image of 30 Doradus, with the {\it Spitzer} Space Telescope 8-$\mu$m IRAC band in red, H$\alpha$ in green, and soft X-rays (0.5--2.0 keV) in blue (details of these data are given in $\S$2). Large and small-scale structures are evident, from thin ionized gas and dust filaments of arcsecond widths to cavities a few arcminutes across filled with hot X-ray gas. The warm ionized gas has several shell-like structures, and many of these are expanding with high velocities ($\sim$100--300 km s$^{-1}$; \citealt{chu94}), suggesting that past supernova explosions have occurred in the region. In addition to a large ionized gas mass ($\sim 8 \times 10^{5} M_{\sun}$; \citealt{kenn84}), the 30 Doradus nebula also has $\sim 10^{6} M_{\sun}$ of CO \citep{johansson}. The CO(1--0) maps of 30 Doradus have revealed 33 molecular cloud complexes in the HII region, and in particular, two elongated clouds of CO mass $\sim 4 \times 10^{5} M_{\sun}$ that form a ``ridge'' West and North of R136 (see the CO contours in Figure~\ref{fig:threecolor}). Estimates of the radius $R_{\rm HII}$ of the nebula range from $\sim$110 pc (\citealt{brandl}; using a revised value of $D$ = 50 kpc) to $\sim$185 pc \citep{kenn84}. The nearly factor of two uncertainty in $R_{\rm HII}$ arises from the complex shape that precludes accurate determination of the radius. In this paper, we will assume $R_{\rm HII}$ = 150 pc.

The properties of 30 Doradus described above demonstrate why this HII region is an ideal candidate for assessing the feedback mechanisms of massive stars. The shear number and energetic output of the OB stars facilitates detailed study of the effects of radiation, winds, supernovae, ionization fronts, etc. Additionally, the proximity of 30 Doradus enables a resolved view of the processes and dynamics associated with starburst activity that was common in the early Universe (e.g., \citealt{meurer,shapley}). Indeed, the relatively instantaneous formation of the concentrated massive stars in R136 make 30 Doradus a ``mini-starburst'' \citep{mini}. 

\section{Data}

We analyzed images of 30 Doradus at several wavelengths. A brief description of these data is given below. 

\subsection{Optical}

We compiled optical photometric data on 30 Doradus from three separate observational programs. For the central 35\arcsec $\times$ 35\arcsec around R136 (with right ascension $\alpha =$ 05$^{\rm h}38^{\rm m}45.5^{\rm s}$ and declination $-$69$^{\circ}$06$^{\prime}$02.7$^{\prime\prime}$), we utilize the photometric results of \cite{mal94} from Hubble Space Telescope Planetary Camera observations. These authors identified over 800 stars within this area and obtained a bolometric luminosity $L_{{\rm bol}} = 7.8 \times 10^{7} L_{\sun}$ for their sources.

At larger distances from R136 out to a few arcminutes, we employ the UBV photometric data of \cite{parker1}. The optical images of 30 Doradus from \cite{parker1} were obtained at the 0.9-m telescope at Cerro Tololo Inter-American Observatory (CTIO), with a field of view of 2.6\arcmin $\times$ 4.1\arcmin\ and 0.49\arcsec pixel$^{-1}$. We followed the analyses of \cite{parker2} to convert their measured apparent UBV magnitudes to absolute bolometric magnitudes. 

For the area outside the field of \cite{parker1}, we use the UBV data of \cite{selman05}. These observations were taken with the Wide Field Imager on the MPG/ESO 2.2-m telescope at La Silla, out to half a degree away from R136 with 0.238\arcsec pixel$^{-1}$. Thus, the three datasets combined provide full coverage of 30 Doradus in the U, B, and V bands. 

\begin{figure}
\includegraphics[width=\columnwidth]{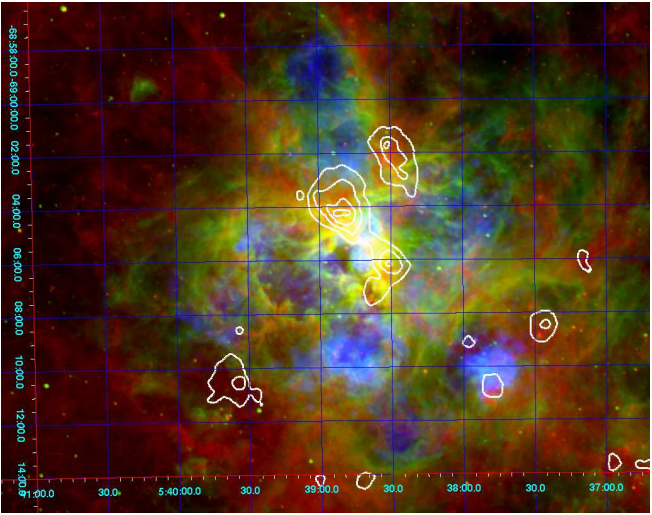}
\caption{Three-color image of 30 Dor: MIPS 8$\mu$m (red), H$\alpha$ (green), and 0.5-8 keV X-rays (blue). White contours show the $^{12}$CO(1-0) emission (Johansson et al. 1998) in the region. Both large- and small-scale structures are evident. North is up, East is left.}
\label{fig:threecolor}
\end{figure}

To illustrate the HII region structure, we show the H$\alpha$ emission of 30 Doradus in Figure~\ref{fig:threecolor}. This narrow-band image (at 6563$\mbox{\AA}$, with 30$\mbox{\AA}$ full-width half max) was taken with the University of Michigan/CTIO 61-cm Curtis Schmidt Telescope at CTIO as part of the Magellanic Cloud Emission Line Survey (MCELS: \citealt{smith}). The total integration time was 600 s, and the reduced image has a resolution of 2\arcsec pixel$^{-1}$.

\subsection{Infrared}

Infrared images of 30 Doradus were obtained through the {\it Spitzer} Space Telescope Legacy project Surveying the Agents of Galaxy Evolution (SAGE: \citealt{meixner}) of the LMC. The survey covered an area of $\sim$7 $\times$ 7 degrees of the LMC with the Infrared Array Camera (IRAC; \citealt{fazio}) and the Multiband Imaging Photometer (MIPS; \citealt{rieke}). Images were taken in all bands of IRAC (3.6, 4.5, 5.8, and 7.9 $\mu$m) and of MIPS (24, 70, and 160 $\mu$m) at two epochs in 2005. For our analyses, we used the combined mosaics of both epochs with 1.2\arcsec pixel$^{-1}$ in the 3.6 and 7.9 $\mu$m IRAC images and 2.49\arcsec pixel$^{-1}$ and 4.8\arcsec pixels$^{-1}$ in the MIPS 24 $\mu$m and 70 $\mu$m, respectively. 

\subsection{Radio}

30 Doradus was observed with the Australian Telescope Compact Array (ATCA) as part of a 4.8-GHz and 8.64-GHz survey of the LMC \citep{dickel}. This program used two array configurations that provided 19 antenna spacings, and these ATCA observations were combined with the Parkes 64-m telescope data of \cite{haynes} to account for extended structure missed by the interferometric observations. For our analyses, we utilized the resulting ATCA$+$Parkes 8.64 GHz (3.5-cm) image of 30 Doradus, which had a Gaussian beam of FWHM 22\arcsec\ and an average rms noise level of 0.5 mJy beam$^{-1}$. We note that higher-resolution ATCA observations of 30 Doradus have been taken by \cite{lazendic}, but we have opted to use the ATCA$+$Parkes image of \cite{dickel} as the latter is more sensitive to the low surface-brightness outskirts of 30 Doradus.

\subsection{X-ray}

\begin{figure}
\includegraphics[width=\columnwidth]{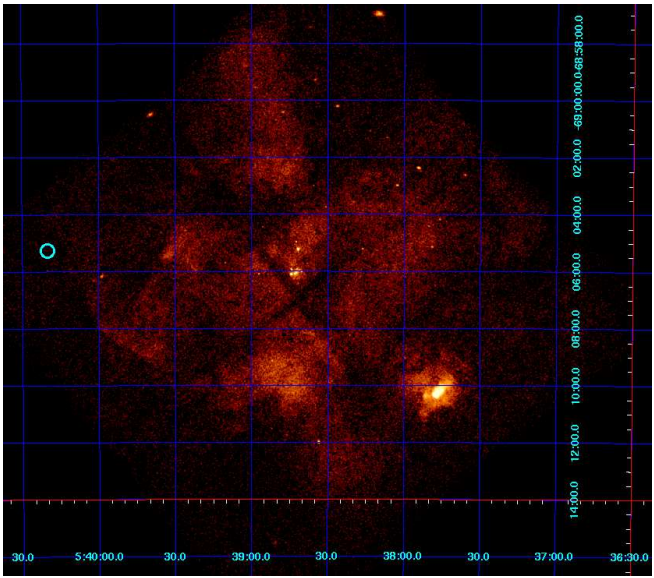}
\caption{{\it Chandra} ACIS X-ray soft band (0.5--2.0 keV) image of 30 Doradus. The image was binned by a factor of four so structures are visually apparent. The cyan circle $\approx$ 2\arcmin\ East of 30 Doradus is the area where background spectra were extracted. North is up, East is left.}
\label{fig:xray}
\end{figure}

30 Doradus was observed using the {\it Chandra} Advanced CCD Imaging Spectrometer (ACIS) in 2006 January for $\approx$94 ks total (ObsIDs 5906 [13 ks], 7263 [43 ks], 7264 [38 ks]; PI: L. Townsley) in the Timed-Exposure VFaint Mode. The spatial resolution of the {\it Chandra} ACIS images is 0.492\arcsec\ pixel$^{-1}$. Data reduction and analysis was performed using the {\it Chandra} Interactive Analysis of Observations ({\sc ciao}) Version 4.1. We followed the {\sc ciao} data preparation thread to reprocess the Level 2 X-ray data and merge the three observations together. Figure~\ref{fig:xray} shows the resulting soft X-ray band (0.5--2.0 keV) image following these analyses. Seventy-four point sources were identified in the reprocessed images using the {\sc ciao} command {\it wavdetect} (a source detection algorithm using wavelet analysis; \citealt{f02}); we excluded the identified point sources in our spectral analyses. 

\begin{figure}
\includegraphics[width=1.00\columnwidth]{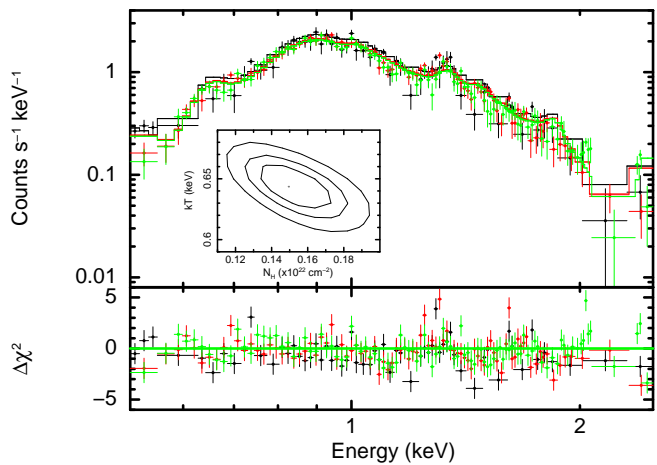}
\caption{The global X-ray spectra from the three ACIS observations of 30 Doradus (ObsID 5906 in black, ObsID 7263 in red, and ObsID 7264 in green), with the best-fit models overplotted. The inset is the 68\%, 90\%, and 99\% confidence contours for the column density $N_{\rm H}$ and the temperature $kT_{\rm X}$. The bottom panel gives the residuals between the data and the model in terms of $\chi^{2}$. We find a best-fit $N_{\rm H} = 1.5_{-0.2}^{+0.3} \times 10^{21}$ cm$^{-2}$ and $kT_{\rm X} = 0.64_{-0.02}^{+0.03}$ keV.}
\label{fig:Xspectra}
\end{figure}

To produce a global X-ray spectrum of 30 Doradus, we extracted {\it Chandra} spectra using the {\sc ciao} command {\it specextract}. Background spectra were also produced from a circular region of radius $\approx$15\arcsec\ that is $\approx$2\arcmin\ East of 30 Doradus, and these were subtracted from the source spectra. Additionally, we removed the counts of the 74 point sources identified above. The resulting spectra were modeled simultaneously as an absorbed, variable-abundance plasma in collisional ionization equilibrium (XSPEC model components {\it phabs} and {\it vmekal}) in XSPEC Version 12.4.0. Figure~\ref{fig:Xspectra} gives the spectra with the best-fit model (with $\chi^2 = 619$ with $396$ degrees of freedom [d.o.f.]) overplotted. We found a best-fit absorbing column density of $N_{\rm H} = 1.5_{-0.2}^{+0.3} \times 10^{21}$ cm$^{-2}$ and an X-ray gas temperature of $kT_{\rm X} = 0.64_{-0.02}^{+0.03}$ keV. The absorption-corrected soft-band (0.5--2.0 keV) luminosity of the diffuse emission is $L_{\rm X} = 4.5 \times 10^{36}$ erg s$^{-1}$.

Previous {\it Chandra} X-ray analysis of 30 Doradus was reported by \cite{townsley1,townsley2} for a different set of observations (ObsIDs 22 and 62520) totalling $\sim$24 ks. By fitting the X-ray spectra of many diffuse regions across 30 Doradus, they found best-fit absorbing columns of $N_{\rm H} = 1-6 \times 10^{21}$ cm$^{-2}$, temperatures of $kT_{\rm X} \sim$ 0.3--0.9 keV, and absorption-corrected luminosities (0.5--2.0 keV) of log$L_{\rm X} =$ 34.2--37.0 erg s$^{-1}$. Thus, our values are fairly consistent with those of Townsley et al.

\section{Methodology}

\begin{figure}
\includegraphics[width=\columnwidth]{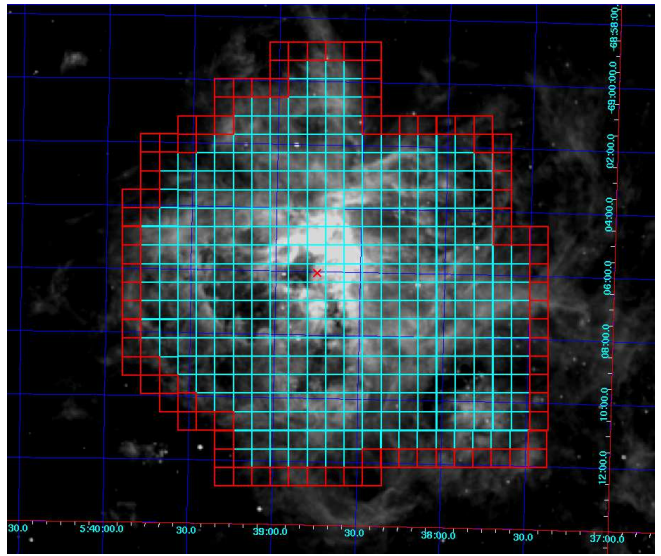}
\caption{H$\alpha$ image (from MCELS; Smith et al. 1998) with the 441 regions we analyzed overplotted. Red squares denote those included in the hot gas leakage analysis of $\S$5.1. The red X marks the center of R136.}
\label{fig:regions}
\end{figure}

To assess how feedback varies spatially across 30 Doradus, we separate the source into 441 regions (see Figure~\ref{fig:regions}). The area of the individual regions was selected to ensure sufficient signal-to-noise across the analyzed wavebands; we chose the width of the regions (35\arcsec $\approx$ 8 pc on a side, at a distance $D =$ 50 kpc to the LMC) to match the HST PC image of R136 \citep{mal94}, so that we could use their $L_{\rm bol}$ value and not have to resolve the individual point sources in the crowded R136 cluster. The number and position of our 441 regions  was determined by the field-of-view and orientation of the 3-cm radio and {\it Chandra} X-ray data. Figure~\ref{fig:regions} shows the H-$\alpha$ image with all the resulting regions overplotted.

To ascertain the dynamical importance of the feedback processes, we compute the pressures for each region using the methods and relations described below. Since protostellar outflows are only important dynamically in low-mass star clusters \citep{matzner07}, we do not expect them to play a big role in 30 Doradus, and we will not consider them in the rest of the text. 

\subsection{Direct Radiation Pressure} \label{sec:Prad}

The light output by stars produces a direct radiation pressure that is associated with the photons' energy and momentum. The resulting radiation pressure $P_{{\rm dir}}$ at some position within the HII region is related to the bolometric luminosity of each star $L_{\rm bol}$ and the distance $r$ the light traveled to reach that point:

\begin{equation}
P_{{\rm dir}}= \sum \frac{L_{{\rm bol}}}{4 \pi r^{2} c}
\label{eq:Pdir}
\end{equation}

\noindent
where the summation is over all the stars in the field. In $\S$5.2, we describe an alternative definition of radiation pressure and compare the results from each case.

The above relation assumes that the stellar radiation are not attenuated by dust. In $\S$~\ref{sec:PIR}, we calculate separately the radiation pressure associated with the light absorbed by dust using {\it Spitzer} IR photometry. Given that our results show that $P_{\rm dir} \gg P_{\rm IR}$ generally (see $\S$~\ref{sec:results}), the assumption that the emitted $L_{\rm bol}$ is unattenuated seems reasonable.

In order to obtain the bolometric luminosities of the massive stars in 30 Doradus, we utilize the UBV photometric data described in $\S$2.1. To simplify the calculation, we assume the bolometric luminosity of R136 obtained by \cite{mal94} originates from the point in the middle of the central region marked with the red X in Fig.~\ref{fig:regions}. For the stars located outside R136 within a few arcminutes, the \cite{parker1} catalog only includes the apparent UBV magnitudes and colors. Therefore, we follow the procedure outlined by \cite{parker2} to obtain absolute bolometric magnitudes of the 1264 stars in the \cite{parker1} catalog that are not in the \cite{selman05} sample. For the 7697 stars in the \cite{selman05} catalog that lie outside the field of the \cite{parker1} data, we use their published values for the stars' absolute bolometric magnitudes. 

Thus, in total, we calculate the bolometric luminosities $L_{\rm bol}$ of the R136 cluster and 8961 other stars in 30 Doradus. For each of the 441 regions, we sum these 8962 terms in Equation~\ref{eq:Pdir}, where $r$ corresponds to the projected distance from the 8962 stars' positions to the region center. In this manner, we compute the radiation presure 'felt' by the 441 regions from all of the starlight in 30 Doradus.

Since the stars are viewed in projection, the actual distance $r$ to a star from the R136 center is observed as a projected distance $\psi$. Therefore, we calculate the direct radiation pressure for two scenarios: one case assuming the stars lie in the same plane ($P_{\rm dir}$) and another case where we attempt to ``deproject'' the stars positions ($P_{\rm dir,3D}$). Appendix A outlines the procedure we utilize to obtain the deprojected bolometric luminosity of the stars as a function of $r$ and compares $P_{\rm dir}$ with $P_{\rm dir,3D}$. We find that $P_{\rm dir,3D}$ is 10--60\% less than $P_{\rm dir}$ at radii $\ls$20 pc from 30 Doradus, and the fractional difference between $P_{\rm dir,3D}$ and $P_{\rm dir}$ at larger radii is much less (0.1--3.0\%). As these differences do not affect the conclusions of this paper, we will only consider $P_{\rm dir}$ for the rest of our analyses.

\subsection{Dust-Processed Radiation Pressure} \label{sec:PIR}

The stars' radiation will be processed by the nearby dust in the region, and an associated pressure is exerted by the resulting infrared radiation field, $P_{\rm IR}$. This pressure component could become dynamically important if the expanding HII shell is optically thick to the IR light, effectively trapping the radiation inside the HII shell \citep{km09}. The pressure of the dust-processed radiation field $P_{{\rm IR}}$ can be determined by the energy density of the radiation field {\it absorbed} by the dust, $u_{\nu}$ (i.e., we assume steady state), 

\begin{equation}
P_{{\rm IR}} = \frac{1}{3} u_{\nu}, 
\end{equation}

To find $u_{\nu}$ in each of our 441 regions, we measure their flux densities $F_{\nu}$ in the IRAC and MIPS images and compare them to the predictions of the dust models of \citealt{dl07} (DL07 hereafter). The DL07 models show how the IR emission spectral-energy distribution varies depending on the dust content and the intensity of radiation heating the dust. DL07 assume a mixture of carbonaceous grains and amorphous silicate grains that have a size distribution that reproduces the wavelength-dependent extinction in the local Milky Way (see \citealt{wd01}). One component of this dust mixture is polycyclic aromatic hydrocarbons (PAHs), small-sized carbonaceous grains that produce strong emission features at $\sim$3--19 $\mu$m observed in many galaxies. 

Since the infrared emission from dust is relatively insensitive to the spectrum of the incident photons with $h \nu <$ 13.6 eV, DL07 adopts the spectrum of the local-neighborhood ISM. Then, $u_{\nu}$ is given by 

\begin{equation}
u_{\nu} = U u_{\nu}^{{\rm IRSF}}
\label{eq:u}
\end{equation}

\noindent
where $U$ is a dimensionless scale factor and  $u_{\nu}^{{\rm IRSF}}$ is the energy density of the $h \nu < 13.6$ eV photons in the local ISM, 8.65 $\times$ 10$^{-13}$ erg cm$^{-3}$ \citep{mathis}. We assume that each region is exposed to a single radiation field because the starlight heating the dust comes largely from NGC 2070. In DL07 parameters, this case corresponds to $U_{\rm min}=U_{\rm max}$ and $\gamma$=0, where $(1-\gamma)$ is the fraction of the dust mass exposed to the radiation. 

For our analyses, we measure the average flux densities $F_{\nu}$ at 8, 24, and 70 $\mu$m wavelengths for the 441 regions. We do not consider the 160 $\mu$m band because its flux density relative to the 70 $\mu$m is much higher than is consistent with the DL07 models. We suspect that the 160 $\mu$m flux is from cold dust that is not associated with 30 Doradus but is in the sight line to the HII region. 

To ensure we are measuring the 8 and 24 $\mu$m flux densities only from dust and not starlight, we remove the starlight contribution at these wavelengths based on the 3.6$\mu$m flux density (which is almost entirely from starlight):

\begin{eqnarray}
F_{\nu}^{\rm{ns}}(8 \mu m) & = & F_{\nu}(8 \mu m) - 0.232F_{\nu}(3.6 \mu m) \\
F_{\nu}^{\rm{ns}}(24 \mu m) & = & F_{\nu}(24 \mu m) - 0.032F_{\nu}(3.6 \mu m)
\end{eqnarray}
 
\noindent
where the left-hand sides are the non-stellar flux at the respective wavelengths. The coefficients 0.232 and 0.032 are given in \cite{helou}. 

\begin{figure}
\includegraphics[width=\columnwidth]{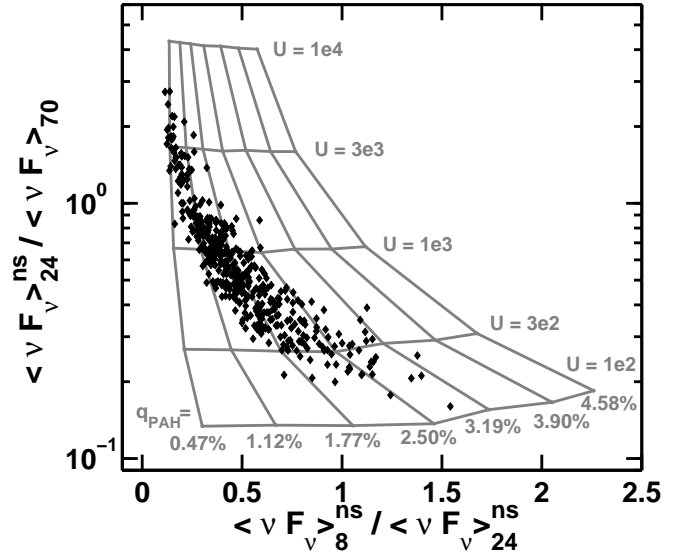}
\caption{Measured IR flux ratios for the 441 regions (black diamonds) and the predicted flux ratios for different PAH mass fractions $q_{{\rm PAH}}$ and scaling $U$ of the energy density of the dust-processed radiation field (Equation~\ref{eq:u}) from DL07. We interpolate the grid of predicted flux ratios to obtain $q_{\rm PAH}$ and $U$ for each region.}
\label{fig:models}
\end{figure} 

To account for the different spatial resolutions of the IR images, we convolved the 3.6, 8, and 24 $\mu$m images with kernels to match the point-spread function of the 70 $\mu$m image. For this analysis, we employed the convolution kernels and method described in $\S$2.3 of \cite{gordon}.

Figure \ref{fig:models} shows the resulting average IR flux ratios, $\langle \nu F_{\nu} \rangle_{{\rm 24}}^{{\rm ns}}/ \langle \nu F_{\nu} \rangle_{{\rm 70}}$ versus $\langle \nu F_{\nu} \rangle_{{\rm 8}}^{{\rm ns}}/ \langle \nu F_{\nu} \rangle_{{\rm 24}}^{{\rm ns}}$, of our regions. Overplotted are the DL07 model predictions for given values of $q_{{\rm PAH}}$, the fraction of dust mass in PAHs, and $U$. Errors in our flux ratios are $\approx$2.8\% from a $\approx$2\% uncertainty in the {\it Spitzer} photometry.

\begin{figure}
\includegraphics[width=\columnwidth]{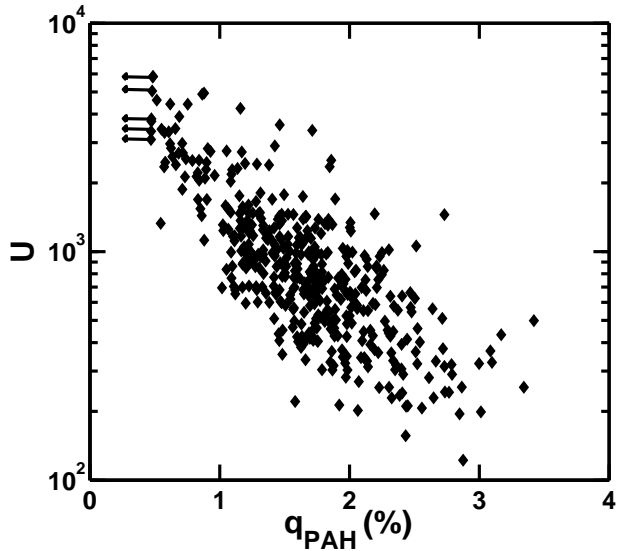}
\caption{Plot of $U$ versus PAH fraction $q_{PAH}$.  Arrows reflect upper-limits in the $q_{\rm PAH}$ values, corresponding to the points outside the grid in Fig.~\ref{fig:models}.}
\label{fig:UvsPAH}
\end{figure}

We interpolate the $U$-$q_{{\rm PAH}}$ grid using Delaunay triangulation, a technique appropriate for a non-uniform grid, to find the $U$ and $q_{{\rm PAH}}$ values for our regions. Figure~\ref{fig:UvsPAH} plots the interpolated values of $U$ versus $q_{{\rm PAH}}$. Since the points with the smallest $\langle \nu F_{\nu} \rangle_{{\rm 8}}^{{\rm ns}}/ \langle \nu F_{\nu} \rangle_{{\rm 24}}^{{\rm ns}}$ values lie to the left of the $U$-$q_{{\rm PAH}}$ grid in Figure~\ref{fig:models}, we are only able to set upper limits of $q_{{\rm PAH}} = $ 0.47\% for them (marked with arrows in Fig.~\ref{fig:UvsPAH}). Thus, these regions produce the ``wall'' of points at $q_{{\rm PAH}} = $ 0.47\% in the $U$ versus $q_{{\rm PAH}}$ plot. 

We find that the PAH fraction spans roughly an order of magnitude, with values up to $q_{{\rm PAH}} =$ 3.76\%. $U$ (and thus $u_{\nu}$) also varies significantly across 30 Doradus, with $U \approx$ 91--7640, corresponding to $u_{\nu} \approx 6.6 \times 10^{-9}-7.9 \times 10^{-11} $ erg cm$^{-3}$. These ranges vary radially, with the largest $U$ and smallest $q_{\rm PAH}$ close to R136. One possible explanation for the $q_{\rm PAH}$ radial dependence is that PAHs are destroyed more where the radiation field heating the dust is strong (e.g., \citealt{raja}). This result is consistent with the analyses of \cite{peeters}, who showed that the ratio of PAH to far-IR (dust continuum) emission in Galactic HII regions is inversely correlated with the intensity of the UV field absorbed by the dust. 
 
We utilize the interpolated $U$ values and Equation~\ref{eq:u} to obtain the energy density $u_{\nu}$, and thus the pressure, of the dust-processed radiation field in the 441 regions. 

\subsection{Warm Ionized Gas Pressure}

\begin{figure}
\includegraphics[width=\columnwidth]{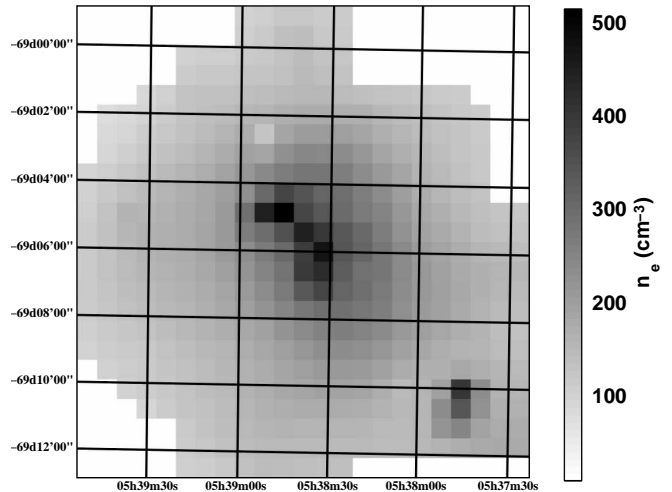}
\caption{Map of electron density $n_{{\rm e}}$ of the warm gas across 30 Doradus, calculated using the bremsstralung flux observed in the 3-cm ATCA data. We find $n_{\rm e}$ is fairly uniform, with a general range of $n_{\rm e} \approx 100-500$ cm$^{-3}$.}
\label{fig:nHII}
\end{figure}

Next, we consider the pressure associated with both the warm HII gas and the hot X-ray gas. The warm ionized gas pressure is given by the ideal gas law, $P_{{\rm HII}} = (n_{{\rm e}} + n_{\rm H} + n_{\rm He}) kT_{{\rm HII}}$, where $n_{{\rm e}}$, $n_{\rm H}$, and $n_{\rm He}$ are the electron, hydrogen, and helium number densities, respectively, and $T_{{\rm HII}}$ is temperature of the HII gas, which we assume to be the same for electrons and ions. If helium is singly ionized, then $n_{{\rm e}} + n_{\rm H} + n_{\rm He} \approx 2 n_{\rm e}$. The temperature of the HII gas in 30 Doradus is fairly homogeneous, with $T_{\rm HII} = 10270\pm140$ K, based on the measurement of [O {\sc iii}] $(\lambda 4959 +\lambda 5007)/\lambda 4363$ across 135 positions in the nebula \citep{krabbe}; here, we adopt $T_{\rm HII}$ = 10$^{4}$ K. Since $T_{\rm HII}$ is so uniform, the warm gas pressure is determined by the electron number density $n_{\rm e}$. We estimate $n_{{\rm e}}$ from the average flux density $F_{\nu}$ of the free-free radio emission in each region (Eq.~5.14b, Rybicki \& Lightman 1979): 

\begin{equation}
n_{\rm e} = \bigg( \frac{6.8 \times 10^{38} 4 \pi D^2 F_{\nu} T_{{\rm HII}}^{1/2}}{\bar{g}_{\rm ff} V} \bigg)^{1/2},
\label{eq:ne}
\end{equation}

\noindent
where we have set the Gaunt factor $\bar{g}_{\rm ff} = 1.2$. In the above relation, $D$ is the distance to 30 Doradus (assumed to be $D$ = 50 kpc) and $V$ is the integrated volume of our regions. For $V$, we assume a radius of the HII region $R$=150 pc, and we calculate the volume by multiplying the area of our region squares by the path length through the sphere at the region's position. We measure $F_{\nu}$ of our regions in the 3.5-cm ATCA$+$Parkes image, since bremsstrahlung dominates at that wavelength. Figure~\ref{fig:nHII} shows the resulting map of $n_{\rm e}$ from these analyses. We find that the central few arcminute area of 30 Doradus has elevated electron densities, with values $n_{\rm e} \approx 200-500$ cm$^{-3}$; the location of these large electron densities corresponds to the two molecular clouds that form the ``ridge'' in the center of the nebula \citep{johansson}. The area outside the central $n_{\rm e}$ enhancement has relatively uniform electron density, with $n_{\rm e} \approx$ 100--200 cm$^{-3}$. In the Southwest of 30 Doradus where the supernova remnant N157B is located, we obtain elevated 3.5-cm flux densities, possibly because of a non-thermal contribution from that source. Therefore, the actual $n_{\rm e}$ may be lower than the values we find in that region.

Our warm gas electron densities are similar to the values obtained by \cite{remy} using {\it Spitzer}'s Infrared Spectrograph. These authors used the ratio [S {\sc iii}] $\lambda$18.7 $\mu$m/[S {\sc iii}]  $\lambda$33.4 $\mu$m to map $n_{\rm e}$ across the central $\approx$2\arcmin\ of 30 Doradus. They also find enhancements in $n_{\rm e}$ along the ``ridge''. 

\subsection{Hot Gas Pressure}

The hot X-ray gas arises from shock heating by stellar winds and supernovae, and the associated hot gas pressure is given by the relation $P_{{\rm X}} = 1.9 n_{{\rm X}} k T_{{\rm X}}$, where $n_{{\rm X}}$ and $T_{{\rm X}}$ are the electron number density and temperature of the X-ray gas, respectively. The factor of 1.9 arises from the assumption that He is fully ionized and that the He mass fraction is 0.3. As in our warm ionized gas calculation, we assume the electrons and ions have reached equipartition, so a single temperature describes both. Since the hot gas can exist over a range of $T_{\rm X}$, we measure both $n_{\rm X}$ and $T_{\rm X}$ by modeling the bremsstrahlung spectrum at X-ray wavelengths. 

From the three {\it Chandra} observations, we extracted {\it Chandra} X-ray spectra from each region using the {\sc ciao} command {\it specextract}. Background spectra were also produced from a circular region of radius $\approx$15\arcsec\ that is $\approx$2\arcmin\ East of 30 Doradus (the cyan circle in Figure~\ref{fig:xray}), and these were subtracted from the source spectra. Resulting spectra were fit using XSPEC Version 12.4.0. Data were grouped such that a minimum of five counts were in each energy bin, and spectra from the three ACIS observations of a given region were fit simultaneously to improve statistics (i.e., they were fit jointly, with more weight given to the data from the longer integrations). Around the edges of the HII region, the X-ray signal is weaker, so we combined adjacent regions to achieve sufficient counts for an accurate fit. 

Spectra were modeled as an absorbed hot diffuse gas in collisional ionization equilibrium using the XSPEC components {\it phabs} and {\it mekal} \citep{m85,m86,l95}. In this fit, we assumed a metallicity $Z \sim 0.5 Z_{\sun}$, the value measured in HII regions in the LMC \citep{kd98}. For regions in the southwest of 30 Doradus with strong emission from the supernova remnant N157B, we added a powerlaw component to account for the non-thermal emission from the SNR. We obtained good fits statistically, with reduced chi-squared values of 0.80--1.30 with 60--300 d.o.f. If a region's fit had reduced chi squared values outside this range or less than 60 d.o.f. we combined its spectra with those of adjacent regions to increase signal. The latter criterion was selected since we found generally that the shape of the bremsstrahlung continuum was not discernable with less than 60 d.o.f. 

From our fits, we can estimate the electron number density $n_{{\rm X}}$ of each region based on the emission measure $EM$ of our models. Emission measure is defined as $EM = \int n_{{\rm X}}^{2} dV$. Thus, 

\begin{equation}
n_{{\rm X}} = \bigg( \frac{EM}{V} \bigg)^{1/2} 
\label{eq:nX}
\end{equation}

\noindent
where $V$ is the integrated volume of our region (the same as used in Eq.~\ref{eq:ne}). Since we are interested in the contribution of the X-ray pressure to the global dynamics, we have divided the emission measure $EM$ by the integrated volume of a region $V$ in calculating $n_{\rm X}$, rather than the volume occupied by the hot gas. In the former case, the density $n_{\rm X}$ goes as the filling factor $f^{-1/2}$; in the latter scenario, $n_{\rm X} \propto f \cdot f^{-1/2} = f^{1/2}$. If the filling factor of the hot gas is small, the thermal pressure of the bubbles may be high internally; however, the hot gas would be insignificant dynamically because it occupies a negligible volume and thus exerts little pressure on the material that bounds the HII region. Therefore, for our purposes of assessing the dynamical role of the hot gas, it is appropriate to use the integrated volue in calculating $n_{\rm X}$.

\begin{figure}
\includegraphics[width=\columnwidth]{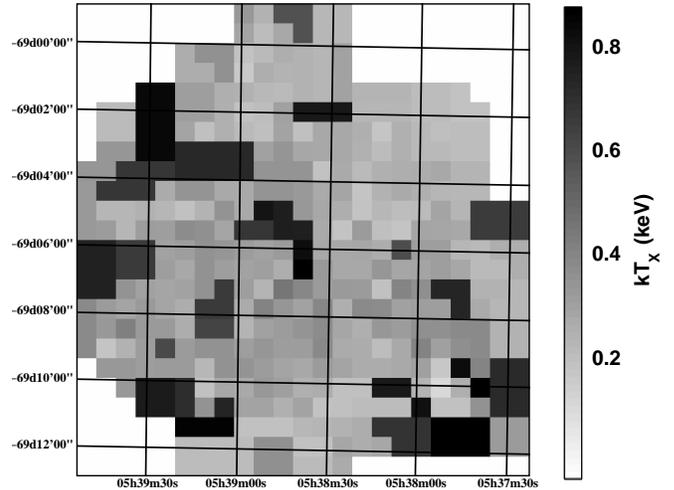}
\caption{Map of X-ray gas temperature $kT_{{\rm X}}$ (in keV) across 30 Doradus. These values were obtained by modeling the {\it Chandra} X-ray spectra from each region.}
\label{fig:kTx}
\end{figure}

\begin{figure}
\includegraphics[width=\columnwidth]{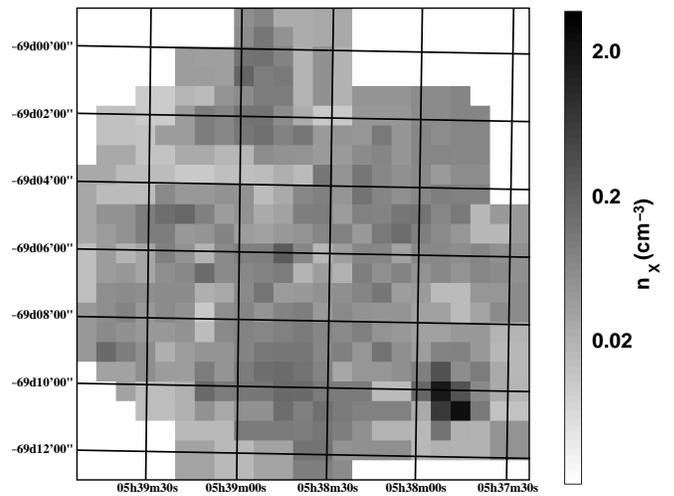}
\caption{Map of the hot gas electron density $n_{{\rm X}}$ (in particles cm$^{-3}$) across 30 Doradus. These values were obtained by modeling the {\it Chandra} X-ray spectra from each region, which output the best-fit emission measure $EM$. We converted $EM$ to $n_{\rm X}$ using Equation 7.}
\label{fig:nX}
\end{figure}

Figure~\ref{fig:kTx} shows a map of the best-fit temperatures $kT_{\rm X}$ from the spectral modeling analyses. The X-ray gas temperatures are elevated in several areas, including in the Southwest (the bottom right of Fig.~\ref{fig:kTx}), where the SNR N157B is located, and at the center near R136. Figure~\ref{fig:nX} gives the map of the hot gas electron density across 30 Doradus from our fits. We find a mean $\langle n_{\rm X} \rangle = 0.12$ cm$^{-3}$. The hot gas electron density is much less than that of the warm gas since many fewer electrons are heated to X-ray emitting temperatures ($\sim10^{7}$ K) than to the moderate temperatures $\sim10^{4}$ K of the warm gas. 

\section{Results} \label{sec:results}

\begin{figure}
\includegraphics[width=\columnwidth]{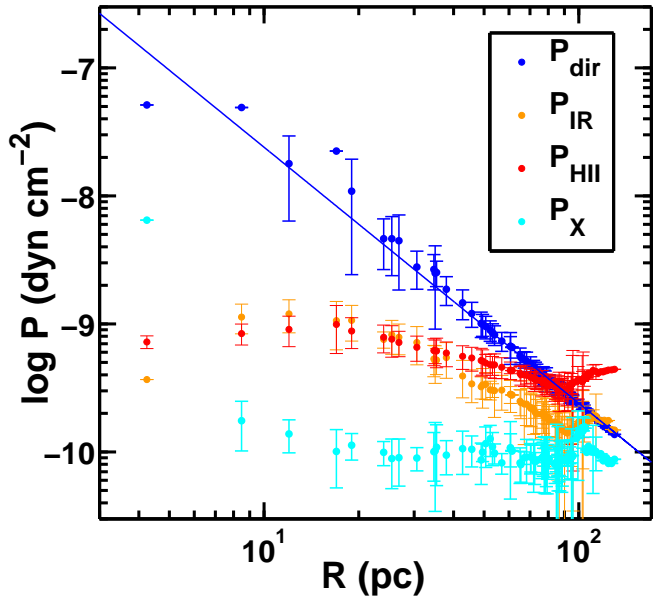}
\caption{All pressures versus radius from the center of R136. Regions with similar radii (defined as radii within 10\% fractionally of each other) are binned to simplify the plot and make trends more readily apparent, and bars reflect the 1-$\sigma$ standard deviations in the pressures at the given radii. Generally, $P_{\rm dir}$ dominates at radii $\ls$75 pc and follows a $P_{\rm dir} \propto r^{-2}$ relation (the blue solid line), whereas $P_{\rm HII}$ dominates at larger distances from R136. $P_{\rm IR}$ and $P_{\rm X}$ do not appear to contribute significantly.}
\label{fig:PvsR}
\end{figure}

Following the multiwavelength analyses and methodology of $\S$3, we calculate the pressures associated with the direct stellar radiation pressure $P_{\rm dir}$, the dust-processed radiation pressure $P_{\rm IR}$, the warm ionized gas pressure $P_{\rm HII}$, and the hot X-ray gas pressure $P_{\rm X}$. Figure~\ref{fig:PvsR} plots the results as a function of distance from the center of R136; data of similar radii (defined as radii within 10\% fractionally of each other) are binned to simplify the plot and to make trends more readily apparent. By comparing the radial trends of the different components, we find that $P_{\rm dir}$ dominates at distances $\ls$75 pc from R136, while $P_{\rm HII}$ dominates at larger radii from R136. Additionally, $P_{\rm IR}$ and $P_{\rm X}$ do not appear to contribute significantly, although they are on the order of $P_{\rm dir}$ at distances $\gs$100 pc from R136. 

As demonstrated by Figure~\ref{fig:PvsR}, we find that $P_{\rm HII} > P_{\rm X}$. The lack of pressure balance between these two components is consistent with our finding (see $\S$5.1) that the X-ray gas does not remain adiabatic and trapped inside the shell. Instead, the hot gas is either leaking out or is mixing with cool gas and suffering rapid radiative losses as a result. In either case, the hot gas is likely to be flowing at a bulk speed comparable to its sound speed, and thus it will not have time to reach pressure equilibrium with the cooler gas that surrounds it before escaping the HII region. Alternatively, it may be that pressure balance is established between the warm ionized gas and the ram pressure of the hot gas, whereas we have only measured the thermal pressure. This picture is consistent with the anti-coincidence of the warm and hot gas noted by previous X-ray work (e.g., \citealt{wang99,townsley1}).

\begin{figure*}
\includegraphics[width=1.05\textwidth]{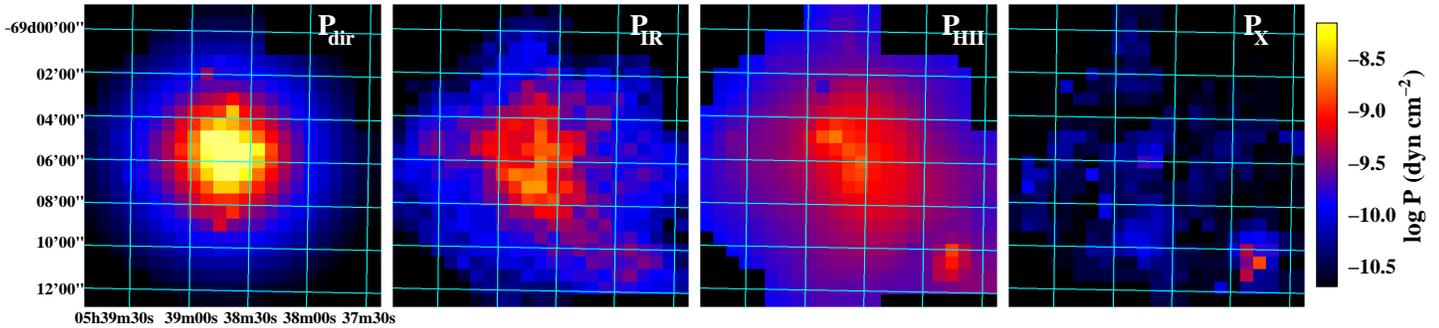}
\caption{Maps of the four pressure components across 30 Doradus. All four are on the same color scale to enable visual comparison. Consistent with Fig.~\ref{fig:PvsR}, $P_{\rm dir}$ dominates in the central few arcminutes, while the $P_{\rm HII}$ dominates at larger distances from R136.}
\label{fig:maps}
\end{figure*}

In Figure~\ref{fig:maps}, we give the maps of the four pressures across 30 Doradus for our 441 regions. $P_{\rm dir}$ has a smooth profile due to its $1/r^{2}$ dependence, while $P_{\rm HII}$ is fairly uniform across 30 Doradus (as expected for a classical HII region). Compared to those components, $P_{\rm IR}$ and $P_{\rm X}$ have more variation throughout the source. Additionally, all the maps have significant enhancements in the central regions near R136; in the cases of $P_{\rm IR}$ and $P_{\rm HII}$, the elevated pressures correspond to the molecular ``ridge'' in 30 Doradus (as seen in the CO contours in Figure~\ref{fig:threecolor}). Additionally, all except $P_{\rm dir}$ have greater pressures in the regions near the SNR N157B (the bottom right of the maps).  

We can utilize the obtained pressures to estimate the total energy of each component. In particular, we measure the total energy density $u$ in a given radius bin of Figure~\ref{fig:PvsR} and multiply by the volume of its shell (where we have set the shell thickness to the difference of the upper and lower bound radius of that bin). We convert pressures $P$ to energy densities $u$ using the relations: $P_{\rm dir} = u_{\rm dir}$, $P_{\rm IR} = \frac{1}{3} u_{\rm IR}$, $P_{\rm HII} = \frac{2}{3} u_{\rm HII}$, and $P_{\rm X} = \frac{2}{3} u_{\rm X}$. Using this approach, we find the following total energies for each component: $E_{\rm dir} = 5.1 \times 10^{53}$ erg, $E_{\rm IR} = 1.7 \times 10^{53}$ erg, $E_{\rm HII} = 2.8 \times 10^{53}$ erg, and $E_{\rm X} = 6.5 \times 10^{52}$ erg. Therefore, the direct and dust-processed radiation fields and the warm ionized gas contribute similarly to the energetics of the region, and every component is $\gs$2 orders of magnitude above the typical kinetic energy of a single SN explosion. 

\section{Discussion}

\subsection{Leakage of the Hot Gas} \label{sec:xray}

As mentioned previously, the X-ray emission in 30 Doradus arises from the shock-heating of gas to temperatures of $\sim10^{7}$ K by stellar winds and supernovae (SNe). These feedback processes eventually carve out large cavities, called bubbles and superbubbles, filled with diffuse X-ray emission. In Fig.~\ref{fig:PvsR}, we demonstrated that the pressure associated with the hot gas $P_{\rm X}$ is comparatively low relative to the other pressure components. Here, we explore the implications of this result in regard to the trapping/leakage of the hot gas. For this discussion, we will consider stellar winds only and ignore the contribution by SNe; this assumption is reasonable given that the mechanical energy of one SN is on the order of the amount injected by winds over a single massive star's lifetime \citep{castor75}. This assumption is valid at the 0.5$Z_{\sun}$ of the LMC: simulations of a $5.5 \times 10^{4} M_{\sun}$ star cluster in Starburst99 \citep{starburst99} showed that the total wind luminosity decreased by roughly a factor of two from the solar to half-solar metallicity case.  

There are several competing theoretical models to account for the X-ray luminosity in bubbles and superbubbles. The models of \cite{castor75} and \cite{weaver77} assumes that the shock-heated gas is completely confined by a cool shell expanding into a uniform-density ISM. An alternative theory proposed by \cite{chev85} ignores the surrounding ISM and employs a steady-state, free-flowing wind. Recently, \cite{hc09} introduced an intermediate model between these two, whereby the ambient ISM is non-uniform. In this case, only some of the hot gas can escape freely through the holes in the shell. 

The fraction of hot gas confined by the shell directly determines the hot gas pressure on the shell as well as the X-ray luminosity within the bubble. If the shell is very porous, the shock-heated gas will escape easily, the wind energy will be lost from the bubble, and the associated pressure and luminosity will be low. By comparison, a more uniform shell will trap the hot gas, retain the wind energy within the bubble, and the corresponding X-ray pressure and luminosity will be much greater. As such, in the latter case, the shocked winds could have a significant role in the dynamics of the HII region. We note that the warm gas is not able to leak similarly because its sound speed is less than the velocities of the shells (20--200 km s$^{-1}$: \citealt{chu94}).

To assess whether the hot gas is trapped inside the shell and is dynamically important, we measure the ratio of the hot gas pressure to the direct radiation pressure, $f_{\rm trap,X} = P_{\rm X}/P_{\rm dir}$, and compare it to what $f_{\rm trap,X}$ would be if all the wind energy was confined. We can calculate the trapped-wind value using the wind-luminosity relation \citep{kud,rep}, which indicates that the momentum flux carried by winds from a star cluster is about half that carried by the radiation field if the cluster samples the entire IMF. Written quantitatively, $0.5 L_{\rm bol}/c = \dot{M_{\rm w}} v_{\rm w}$, where $\dot{M_{\rm w}}$ is the mass flux from the winds that launched at a velocity $v_{\rm w}$. The mechanical energy loss $L_{\rm w}$  of the winds is then given by 

\begin{equation}
L_{\rm w} = \frac{1}{2} \dot{M_{\rm w}} v_{\rm w}^{2} = \frac{L_{\rm bol}^2}{8 \dot{M_{\rm w}} c^2},
\label{eq:Lwind}
\end{equation}

\noindent
and the mechanical energy of the winds is simply $E_{\rm w} = L_{\rm w} t$, where $t$ is the time since the winds were launched. Putting these relations together, the trapped X-ray gas pressure $P_{\rm X,T}$ is

\begin{equation}
P_{\rm X,T} = \frac{2 E_{\rm w}}{3 V_{\rm HII}} = \frac{L_{\rm bol}^2 t}{16 \pi \dot{M_{\rm w}} c^2 R_{\rm HII}^3}, 
\end{equation}

\noindent 
where $V_{\rm HII}$ is the volume of the HII region. 

Given that $P_{\rm dir} = L_{\rm bol}/(4 \pi R_{\rm HII}^2 c)$, then $f_{\rm trap,X}$ is 

\begin{equation}
f_{\rm trap,X} = \frac{L_{\rm bol} t}{4 \dot{M_{\rm w}} c R_{\rm HII}} = \frac{L_{\rm bol}}{4 \dot{M_{\rm w}} c v_{\rm sh}}, 
\label{eq:ftrap}
\end{equation}

\noindent
where we have set $R_{\rm HII}/t = v_{\rm sh}$, the velocity of the expanding shell. Finally, we put $\dot{M_{\rm w}}$ in terms of $L_{\rm bol}$ and $v_{\rm w}$, so that Eq.~\ref{eq:ftrap} reduces to

\begin{equation}
f_{\rm trap,X} = \frac{v_{\rm w}}{2 v_{\rm sh}}.
\end{equation}

We use the above equation to obtain an order-of-magnitude estimate of $f_{\rm trap,X}$ if all the wind energy is confined by the shell. We assume a wind velocity $v_{\rm w} \sim 1000$ km s$^{-1}$ (the escape velocity from a O6 V star; a reasonable order-of-magnitude estimate, since O3 stellar winds are faster and WR winds would be slower than this value). If we set $v_{\rm sh} \sim$ 25 km s$^{-1}$ (the expansion velocity over 30 Doradus given by optical spectroscopy; \citealt{chu94}), then $f_{\rm trap,X} \sim$ 20.

\begin{figure}
\includegraphics[width=\columnwidth]{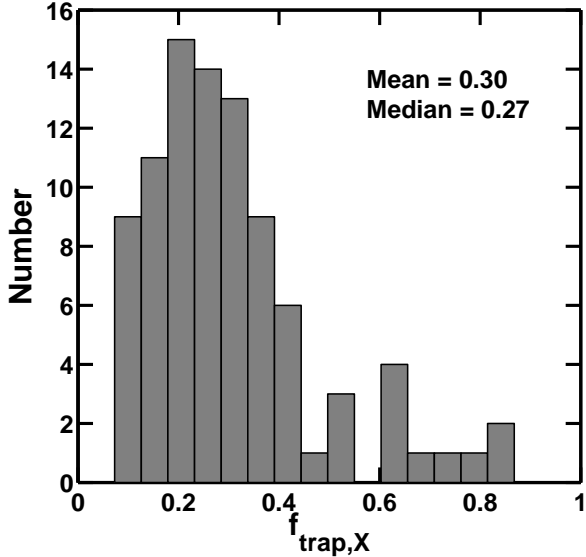}
\caption{Histogram of $f_{\rm trap,X} = P_{\rm X}/P_{\rm dir}$, the ratio of hot gas pressure to the direct radiation pressure, for the regions that are along the rim of our 441 squares in Fig.~\ref{fig:regions}. We find that the mean $f_{\rm trap,X}$ is 0.30 and the median is 0.27, far below the values expected if the hot gas is completely confined in 30 Doradus ($f_{\rm trap,X} \sim $ 20; see text). This result is evidence that the hot gas is leaking out of the shell.}
\label{fig:hist}
\end{figure}

\begin{figure}
\includegraphics[width=\columnwidth]{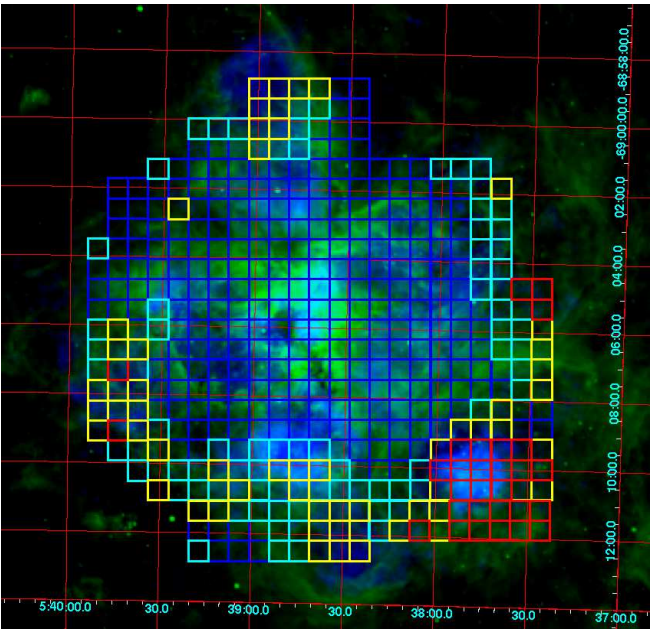}
\caption{H$\alpha$ (green) and soft X-ray (blue) images with the region grid overplotted, where the regions are color-coded based on their $f_{\rm trap,X} = P_{\rm X}/P_{\rm dir}$ values: dark blue = $f_{\rm trap,X} < 0.2$; cyan = $0.2 < f_{\rm trap,X} < 0.3$; yellow = $0.3 < f_{\rm trap,X} < 0.5$; red = $f_{\rm trap,X}>0.5$. The highest values of $f_{\rm trap,X}$ occur near the SNR N157B as well as other areas with strong X-ray emission and weak H$\alpha$ emission.}
\label{fig:ftrapcheck}
\end{figure}

We can compare this $f_{\rm trap,X}$ to our observed values for the regions closest to the shell (the ones along the rim of our 441 squares in Fig.~\ref{fig:regions}); Figure~\ref{fig:hist} shows the histogram of our observed $f_{\rm trap,X}$ values. We find a mean and median $f_{\rm trap,X}$ of 0.30 and 0.27, respectively, for our outermost regions. Over 30 Doradus, the highest values of $f_{\rm trap,X}$ are near the supernova remnant N157B in the southwest corner of 30 Doradus (see Figure~\ref{fig:ftrapcheck}), where hot gas is being generated and has not had time to vent. Other locations where $f_{\rm trap,X}$ is elevated are regions with strong X-ray emission and weak H$\alpha$ emission. Morphologically, these areas could be where the hot gas is blowing out the 30 Doradus shell.

The observed $f_{\rm trap,X}$ values are 1--2 orders of magnitude below what they would be if the wind was fully confined. As a consequence, we find that $P_{\rm X}$ of our regions is too low to be completely trapped in the HII region (the Castor et al. model), and the X-ray gas must be leaking through pores in the shell. This result is consistent with the Harper-Clark \& Murray model of partial confinement of the hot gas, and the weakness of $P_{\rm X}$ relative to $P_{\rm dir}$ suggests the hot gas does not play a significant role in the dynamics of the HII region.

We note here that our rim regions in this analysis are $\sim$70--130 pc from R136, which is less than the estimated radius of $R_{\rm HII} = 110-185$ pc. Therefore, our $f_{\rm trap,X}$ values are lower limits, and the true $f_{\rm trap,X}$ at the shell may be greater by a factor of a few. Nonetheless, the conclusions would remain the same. 

An alternative explanation for the weak X-ray luminosity is that the hot gas mixes with the cool gas, and the hot gas temperature is lowered enough so that the gas can cool efficiently. In that case, the energy is still lost from the system, via radiative cooling instead of the escape of the X-ray emitting material. Far ultraviolet spectroscopy is necessary to determine the level of mixing between the gas components.

\subsection{On the Definition of Radiation Pressure} \label{sec:diffPrads}

In this paper, we have defined the radiation pressure as related to the energy and momentum flux of the light radiated by the stars in 30 Doradus. Alternatively, radiation pressure could be characterized as the force per unit area exerted by the radiation on matter. The two cases produce divergent results regarding the radial dependence of $P_{\rm dir}$. In particular, the former case has large $P_{\rm dir}$ close to the star cluster and a decline in $P_{\rm dir}$ with distance from the center. By contrast, the latter predicts small $P_{\rm dir}$ in the HII-region interior, and $P_{\rm dir}$ becomes significant near the neutral shell where the radiation will be absorbed (see Appendix $\S$B).

Each definition of radiation pressure reveals distinct information about an HII region. When considering the global dynamics of expansion of an HII region, it is necessary to characterize $P_{\rm dir}$ as the energy density of the radiation field, since that definition reflects the total energy and momentum budget available to drive motion. Alternatively, measurement of the force exerted by radiation on matter facilitates a local estimate of the internal density distribution of an HII region. As we are interested principally on the dynamical role of radiation pressure, we have adopted the former definition of $P_{\rm dir}$ in this paper.

\subsection{HII Region Dynamics}

In $\S$\ref{sec:results}, we found that the direct radiation pressure $P_{\rm dir}$ dominates over the ionized gas pressure $P_{\rm HII}$ at radii $\ls$75 pc, implying that the radiation has played a role in the dynamics in 30 Doradus. Significant radiation pressure alters the properties of an HII region (e.g., the density profile: \citealt{draine}) and causes the expansion to proceed differently than in a classical HII region with ionized gas-driven expansion. In particular, \cite{km09} demonstrated that radiation-driven expansion imparts more momentum to the shell, accelerating the expansion at early times relative to that of gas-driven expansion. Indeed, an additional force must have dominated at early times in 30 Doradus since the shell velocity $v_{\rm sh} \sim$ 25 km s$^{-1}$ \citep{chu94} is too fast to have been gas-driven alone because the HII gas sound speed is $c_{\rm s} \approx$ 10 km s$^{-1}$.

To determine the characteristic radius $r_{\rm ch}$ where the HII region shell transitioned from radiation-pressure driven to gas-pressure driven, we can set these pressure terms at the shell equal and solve for $r_{\rm ch}$. Broadly, the pressures at the shell have different dependences with the shell radius $r_{\rm HII}$: $P_{\rm dir} \propto r_{\rm HII}^{-2}$ and $P_{\rm HII} \propto r_{\rm HII}^{-3/2}$. Setting them equal and solving for $r_{\rm ch}$ (Equation 4 in \citealt{km09}, the ``embedded case''), we find

\begin{equation}
r_{\rm ch} = \frac{\alpha_{\rm B}}{12 \pi \phi} \bigg( \frac{\epsilon_{\rm 0}}{2.2 k_{\rm B} T_{\rm HII}} \bigg)^2 f_{\rm trap,tot}^{2} \frac{\psi^2 S}{c^{2}},
\end{equation} 

\noindent
where $\alpha_{\rm B}$ is the case-B recombination coefficient and $\epsilon_{\rm 0} = 13.6$ eV, the photon energy necessary to ionize hydrogen. The dimensionless quantity $\phi$ accounts for dust absorption of ionizing photons and for free electrons from elements besides hydrogen; $\phi=0.73$ if He is singly ionized and 27\% of photons are absorbed by dust (typical for a gas-pressure dominated HII region; \citealt{mw97}). The $f_{\rm trap,tot}$ represents the factor by which radiation pressure is enhanced by trapping energy in the shell through several mechanisms; the trapped hot gas $f_{\rm trap,X}$ calculated in $\S$5.1 is one component that can contribute to $f_{\rm trap,tot}$ (as discussed in $\S$\ref{sec:xray}). Here, we adopt $f_{\rm trap,tot} = 2$, as in \cite{km09}. Lastly, $\psi$ is the ratio of bolometric power to the ionizing power in a cluster; we set $\psi = 3.2$ using the $\langle S \rangle/ \langle M_{*} \rangle$ and the $\langle L \rangle/ \langle M_{*} \rangle$ relations of \cite{mr10}. 

Putting all these terms together, we find $r_{\rm ch} \approx$ 33 pc. Physically, this result means that early in the expansion before it reached a radius of 33 pc, 30 Dor's dynamics could have been radiation-pressure dominated, and it has since become gas-pressure dominated. Alternatively, it is possible that the hot gas pressure dominated at early times and has become weaker as the HII region expands. 

The radiation-driven or hot gas driven expansion at early times in 30 Doradus would have facilitated the expulsion of gas from the central star cluster. In particular, since the warm gas sound speed ($\sim$10 km s$^{-1}$) is less than the escape velocity of R136 ($\sim$20 km s$^{-1}$, given a mass $M=5.5 \times 10^{4} M_{\sun}$ in a radius $R$ = 1 pc; \citealt{hunter95}), an alternative mechanism is necessary to remove the gas and regulate star formation (e.g., \citealt{km09,fall}). We conclude that the radiation pressure or hot gas pressure likely played this role in 30 Doradus, decreasing the available mass to make new stars and slowing star formation in the region.

\smallskip

\section{Summary}

In this paper, we have utilized multi-wavelength (radio, infrared, optical/UV, and X-ray) imaging to assess the role of several stellar feedback mechanisms in the giant HII region 30 Doradus in the LMC. In particular, we have measured observationally the pressures associated with possible sources of energy and momentum to drive the dynamics of the region: the direct radiation from stars, the dust-processed infrared radiation field, the warm ionized gas from massive stars, and the hot gas shock-heated by stellar winds and supernovae. We have exploited the high-resolution images of 30 Doradus to map these pressure components in 441 square regions, with dimensions of 35\arcsec $\times$ 35\arcsec. We have found that the direct radiation pressure from stars dominates at distances less than 75 pc from the central star cluster, whereas the warm ionized gas pressure dominates at larger radii. By contrast, the hot gas pressure and the dust-processed radiation pressure do not contribute significantly, indicating these components are not dynamically important. However, we cannot rule out that the hot gas pressure dominated at early times and has become weaker with the HII region expansion.

We have discussed two implications of our results: the partial confinement of the hot gas and the dynamical role of radiation pressure in 30 Doradus. First, the weakness of the X-ray gas pressure relative to the direct radiation pressure suggests the hot gas is only partially confined and is leaking out of the pores in the HII shell. Secondly, the significant radiation pressure near the star cluster indicates that radiation pressure may have driven the expansion of the HII shell at early times. This result suggests observationally that radiation pressure may be dynamically important in massive star clusters, reinforcing that radiation pressure is a viable mechanism to remove gas from HII regions and to regulate star formation. Indeed, if NGC 2070 was more massive, the radiation pressure could even expel gas at high enough velocities to launch a galactic wind \citep{wind}.

The work presented here is a first step to measure observationally the relative role of stellar feedback mechanisms in star-forming regions. Although we have applied our techniques to one source, 30 Doradus, our methods to extract dynamical information from multi-wavelength images can be applied to other sources as well. Consequently, we plan to perform these analyses on all the HII regions in the LMC with available data to develop a broad observational understanding of these stellar feedback mechanisms and their role in regulating star formation. 

\acknowledgements

We would like to thank Bruce Draine, Xander Tielens, and Fernando Selman for helpful discussions. Also, we would like to thank Jasmina Lazendic-Galloway for generously providing the 3.5-cm ATCA image of 30 Doradus. This work is supported by an AAUW American Dissertation Fellowship (LAL) and by the National Science Foundation through grant NSF-AST0955836. MRK acknowledges support from: an Alfred P. Sloan Fellowship; NASA through ATFP grant NNX09AK31G; NASA as part of the Spitzer Theoretical Research Program, through a contract issued by the JPL; the National Science Foundation through grant AST-0807739. 

\begin{appendix}

\section{Deprojecting the Stars in 30 Dor}

\begin{figure*}
\begin{center}
\includegraphics[width=0.5\columnwidth]{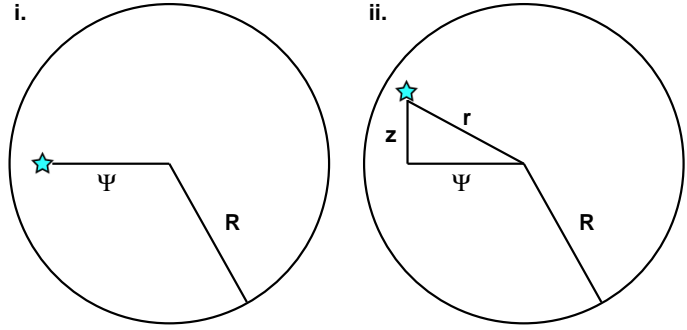}
\end{center}
\caption{Diagram explaining how projection effects may lead to erroneous measurement of the actual distance $r$ of a star to the star cluster center. {\it i}. Face-on view of an HII region of radius $R$. The projected distance from the star to the center is $\psi$. {\it ii}. View from above of the same HII region. In this case, it is apparent that the star does not lie in the midplane of the sphere, and the actual distance is $r = \sqrt{\psi^{2}+z^{2}}$, where $z$ is the line-of-sight distance to the star from the sphere's midplane.}
\label{fig:geo}
\end{figure*}

Since the stars are viewed in projection, the actual distance $r$ to a star from the R136 center is observed as a projected distance $\psi$ (see Figure~\ref{fig:geo}). Therefore, we calculate the direct radiation pressure for two scenarios: one case assuming the stars lie in the same plane (i.e., $r = \psi$; $P_{\rm dir}$) and another case where we attempt to ``deproject'' the stars positions (i.e., $r = \sqrt{\psi^{2}+z^{2}}$, where $z$ is the line-of-sight distance to the star from the sphere's midplane; $P_{\rm dir,3D}$). The direct observable is the projected surface brightness $\mu$ (in units of erg cm$^{-2}$ s$^{-1}$) in an annulus, and it is a function of $\psi$. The luminosity density (in units of erg cm$^{-3}$ s$^{-1}$) $\mathcal{L}(r) = \mathcal{L}(\sqrt{\psi^2+z^2})$ is then related to $\mu(\psi)$ by

\begin{equation}
\mu(\psi) = 2 \int_0^{\sqrt{R^2-\psi^2}} \mathcal{L}(\sqrt{\psi^2+z^2}) dz.
\end{equation}

\noindent
If we put this integral in terms of $r$, we obtain the relation

\begin{equation}
\mu(\psi) = 2 \int_{\psi}^R r(r^2-\psi^2)^{-1/2} \mathcal{L}(r) dr = \int_{R}^{\psi} K(r,\psi) \mathcal{L}(r) dr
\label{eq:Voltra}
\end{equation}

\noindent
where $K(r,\psi) = -2r(r^2-\psi^2)^{-1/2}$.

Equation~\ref{eq:Voltra} is a Volterra equation of the first kind, and we solve for $\mathcal{L}(r)$ explicitly for annuli beginning at some radius $R$ and taking $N$ uniform steps of size $h$ inward to $r_1 = R-hN$. In this case, we selected a radius $R = 200$ pc $\approx$ 825.5\arcsec to ensure the entire nebular volume was included. Additionally, we chose a step size of $h = 1$\arcsec and went inward to $r_{1} = 17.5$\arcsec (so $N$ = 808 steps), the radius of the {\it HST} PC image. 

\begin{figure}
\begin{center}
\includegraphics[width=0.6\columnwidth]{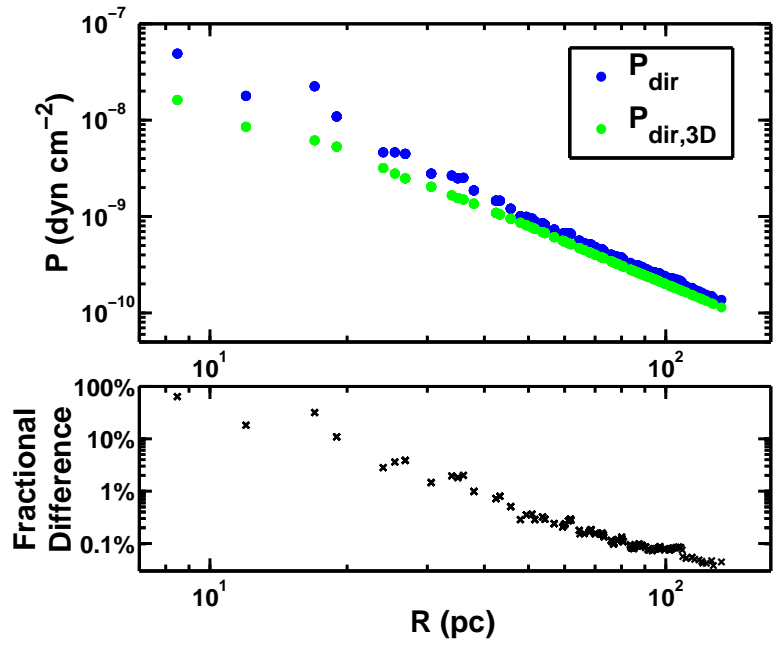}
\end{center}
\caption{Comparison of $P_{\rm dir,3D}$ and $P_{\rm dir}$. {\it Top}: The two pressure components versus distance $R$ from R136. The two have similar radial dependence. {\it Bottom}: The fractional difference $(P_{\rm dir}-P_{\rm dir,3D})/P_{\rm dir}$ for all the points in the top panel. At radii $\gs$20 pc from R136, the fractional difference is small (0.1\%--3.0\%), while at radii $\ls$20 pc, the fractional difference is greater (10\%--60\%).}
\label{fig:Pdircompare}
\end{figure}

Figure~\ref{fig:Pdircompare} (top panel) plots the resulting $P_{\rm dir,3D}$ (and $P_{\rm dir}$ for comparison) versus the distance $R$ from R136. The bottom panel gives the fractional difference $(P_{\rm dir}-P_{\rm dir,3D})/P_{\rm dir}$ for all the points in the top panel. The fractional difference between $P_{\rm dir,3D}$ and $P_{\rm dir}$ is small ($\sim$0.1\%--3.0\%) for regions $\gs$20 pc from R136, and becomes larger ($\sim$10\%--60\% for radii $\ls$20 pc. Despite these greater fractional differences at smaller radii, $P_{\rm dir,3D}$ would still dominate over the other pressure components in Fig.~\ref{fig:PvsR} at distances $\ls$75 pc from R136. Additionally, the small fractional differences at distances $\gs$20 pc confirms that our values of $P_{\rm dir}$ near the shell are accurate. 

We note in the above calculation, we necessarily assumed that the luminosity in a shell is spherically symmetric. The surface brightness in 30 Doradus is not actually symmetric though (since stars and star clusters are distributed non-uniformly around the nebula), so our estimated $P_{\rm dir,3D}$ values are an approximation of the true, deprojected radiation pressure. Nonetheless, the small differences between $P_{\rm dir,3D}$ and  $P_{\rm dir}$ indicate that uncertainty in the star position along the line of sight does not qualitatively affect our results. 

\section{On the Definition of Radiation Pressure}

An alternate definition of radiation pressure than the one used in this paper is to characterize $P_{\rm dir}$ as the force per unit area exerted by the radiation on matter. This case predicts small $P_{\rm dir}$ in the HII-region interior, where the density is small, and $P_{\rm dir}$ only becomes significant near the neutral shell where the radiation is absorbed. To demonstrate this effect, we calculate $f$, the force per unit volume on matter from radiation, as a function of radius in an idealized HII region using Version 08.00 of the photoionization code {\sc cloudy} \citep{ferland}. Assuming photoionization balance, we have

\begin{equation}
\alpha_{\rm B} n_{\rm e} n_{\rm p} = \frac{S(r)}{4 \pi r^2} \sigma_{\rm HI} n_{\rm HI},
\end{equation}

\noindent
where $\alpha_{\rm B} = 2.6 \times 10^{-13}$ cm$^{3}$ s$^{-1}$ is the case-B recombination coefficient of hydrogen at 10$^{4}$ K, $n_{\rm e}$, $n_{\rm p}$, $n_{\rm HI}$ are the electron, proton, and HI number densities, $S$ is the ionizing photon luminosity (in photons s$^{-1}$) passing through a shell at a given radius $r$, and $\sigma_{\rm HI}$ is the H ionization cross-section. 

The force density $f$ is given by

\begin{equation}
f = \frac{\kappa \rho F}{c} = \frac{\kappa_{\rm dust} \rho F}{c} + \frac{\kappa_{\rm HI} \rho F}{c}, 
\end{equation}

\noindent
where $\kappa$ is the opacity per unit mass (from dust $\kappa_{\rm dust}$ or from neutral hydrogen $\kappa_{\rm HI}$), $\rho$ is the local mass density, and $F$ is the total flux. Given $\kappa_{\rm dust} \rho = n_{\rm H} \sigma_{\rm dust}$ (where $n_{\rm H} = n_{\rm HI}+n_{\rm HII}$) and $\kappa_{\rm HI} \rho = n_{\rm HI} \sigma_{\rm HI}$, 

\begin{equation}
f = \frac{S(r) \langle h \nu \rangle}{4 \pi r^2 c} (n_{\rm H} \sigma_{\rm dust} + n_{\rm HI} \sigma_{\rm HI}) = \frac{\alpha_{\rm B} n_{\rm e} n_{\rm p} \langle h \nu \rangle}{c} \bigg(1+ \frac{n_{\rm H}}{n_{\rm HI}} \frac{\sigma_{\rm dust}}{\sigma_{\rm HI}} \bigg).
\label{eq:fbig}
\end{equation}

\noindent
The first term represents the force of ionizing photon absorption by H atoms and the second term is the force of ionizing photons on the dust. Here, $\langle h \nu \rangle$ is the mean energy of the ionizing photons, assuming the force in the radiation field is from ionizing photons only. In the following calculation, we set $\langle h \nu \rangle$ = 15 eV, a value typical of an O star. Similarly, we assume the force on dust is from ionizing photons, and we adopt a dust cross section at 15 eV, $\sigma_{\rm dust} = 1.0 \times 10^{-21}$ cm$^{2}$/H \citep{wd01}. In the above expression, we set $\sigma_{\rm HI} = 6.3 \times 10^{-18}$ cm$^{-2}$.

\begin{figure}
\begin{center}
\includegraphics[width=0.55\columnwidth]{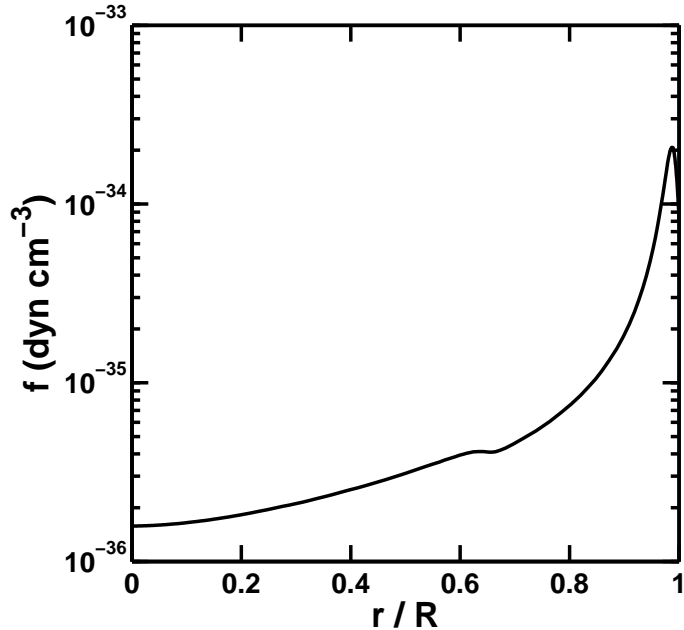}
\end{center}
\caption{Plot of $f$, the force per unit volume on matter from radiation, versus fractional radius $r/R$ of the HII region. The force density $f$ peaks near the HII shell and is several orders of magnitude less interior to the shell.}
\label{fig:fvsr}
\end{figure}

The radial dependence of $f$ comes from the density profiles, $n_{\rm e}(r)$ and $n_{\rm p}(r)$. To estimate these parameters as a function of radius in {\sc cloudy}, we utilize the OSTAR2002 stellar atmosphere model for a metallicity $Z/Z_{\sun} = 0.5$ \citep{lanz} and stellar temperature of $T_{*} = 35000$ K. In our analysis, we set our idealized HII region to have $S = 6.9 \times 10^{49}$ photons s$^{-1}$ and $n_{\rm H} = 10$ cm$^{-3}$. Additionally, we include the effects of interstellar grains in the calculation. 

Figure~\ref{fig:fvsr} plots the resulting $f$ as a function of fractional radius in the HII region. The force per unit volume of the radiation increases many orders of magnitude from the center to the edge of the HII region, with a drastic jump in $f$ at $r/R \gs 0.9$. This plot contrasts Figure~\ref{fig:PvsR}, where $P_{\rm dir}$ falls off as $1/r^{2}$.

\end{appendix}

\bibliographystyle{apj}
\bibliography{ms}

\begin{thebibliography}{91}
\expandafter\ifx\csname natexlab\endcsname\relax\def\natexlab#1{#1}\fi

\bibitem[{{Andrews} \& {Thompson}(2011)}]{andrews}
{Andrews}, B.~H., \& {Thompson}, T.~A. 2011, \apj, 727, 97

\bibitem[{{Ballesteros-Paredes} \& {Hartmann}(2007)}]{ball07}
{Ballesteros-Paredes}, J., \& {Hartmann}, L. 2007, Revista Mexicana de
  Astronomia y Astrofisica, 43, 123

\bibitem[{{Bradley} {et~al.}(2006){Bradley}, {Knapen}, {Beckman}, \&
  {Folkes}}]{bradley}
{Bradley}, T.~R., {Knapen}, J.~H., {Beckman}, J.~E., \& {Folkes}, S.~L. 2006,
  \aap, 459, L13

\bibitem[{{Brandl}(2005)}]{brandl}
{Brandl}, B.~R. 2005, in Astrophysics and Space Science Library, Vol. 329,
  Starbursts: From 30 Doradus to Lyman Break Galaxies, ed. {R.~de Grijs \&
  R.~M.~Gonz{\'a}lez Delgado}, 49--+

\bibitem[{{Castor} {et~al.}(1975){Castor}, {McCray}, \& {Weaver}}]{castor75}
{Castor}, J., {McCray}, R., \& {Weaver}, R. 1975, \apjl, 200, L107

\bibitem[{{Chevalier} \& {Clegg}(1985)}]{chev85}
{Chevalier}, R.~A., \& {Clegg}, A.~W. 1985, \nat, 317, 44

\bibitem[{{Chu} \& {Kennicutt}(1994)}]{chu94}
{Chu}, Y., \& {Kennicutt}, Jr., R.~C. 1994, \apj, 425, 720

\bibitem[{{Cunningham} {et~al.}(2006){Cunningham}, {Frank}, {Quillen}, \&
  {Blackman}}]{cunningham}
{Cunningham}, A.~J., {Frank}, A., {Quillen}, A.~C., \& {Blackman}, E.~G. 2006,
  \apj, 653, 416

\bibitem[{{Dale} {et~al.}(2005){Dale}, {Bonnell}, {Clarke}, \& {Bate}}]{dale}
{Dale}, J.~E., {Bonnell}, I.~A., {Clarke}, C.~J., \& {Bate}, M.~R. 2005,
  \mnras, 358, 291

\bibitem[{{Dickel} {et~al.}(2005){Dickel}, {McIntyre}, {Gruendl}, \&
  {Milne}}]{dickel}
{Dickel}, J.~R., {McIntyre}, V.~J., {Gruendl}, R.~A., \& {Milne}, D.~K. 2005,
  \aj, 129, 790

\bibitem[{{Draine}(2010)}]{draine}
{Draine}, B.~T. 2010, ArXiv e-prints

\bibitem[{{Draine} \& {Li}(2007)}]{dl07}
{Draine}, B.~T., \& {Li}, A. 2007, \apj, 657, 810

\bibitem[{{Elmegreen}(2000)}]{elm00}
{Elmegreen}, B.~G. 2000, \apj, 530, 277

\bibitem[{{Fall} {et~al.}(2010){Fall}, {Krumholz}, \& {Matzner}}]{fall}
{Fall}, S.~M., {Krumholz}, M.~R., \& {Matzner}, C.~D. 2010, \apjl, 710, L142

\bibitem[{{Fazio} {et~al.}(2004){Fazio}, {Hora}, {Allen}, {Ashby}, {Barmby},
  {Deutsch}, {Huang}, {Kleiner}, {Marengo}, {Megeath}, {Melnick}, {Pahre},
  {Patten}, {Polizotti}, {Smith}, {Taylor}, {Wang}, {Willner}, {Hoffmann},
  {Pipher}, {Forrest}, {McMurty}, {McCreight}, {McKelvey}, {McMurray}, {Koch},
  {Moseley}, {Arendt}, {Mentzell}, {Marx}, {Losch}, {Mayman}, {Eichhorn},
  {Krebs}, {Jhabvala}, {Gezari}, {Fixsen}, {Flores}, {Shakoorzadeh}, {Jungo},
  {Hakun}, {Workman}, {Karpati}, {Kichak}, {Whitley}, {Mann}, {Tollestrup},
  {Eisenhardt}, {Stern}, {Gorjian}, {Bhattacharya}, {Carey}, {Nelson},
  {Glaccum}, {Lacy}, {Lowrance}, {Laine}, {Reach}, {Stauffer}, {Surace},
  {Wilson}, {Wright}, {Hoffman}, {Domingo}, \& {Cohen}}]{fazio}
{Fazio}, G.~G., {et~al.} 2004, \apjs, 154, 10

\bibitem[{{Ferland} {et~al.}(1998){Ferland}, {Korista}, {Verner}, {Ferguson},
  {Kingdon}, \& {Verner}}]{ferland}
{Ferland}, G.~J., {Korista}, K.~T., {Verner}, D.~A., {Ferguson}, J.~W.,
  {Kingdon}, J.~B., \& {Verner}, E.~M. 1998, \pasp, 110, 761

\bibitem[{{Freeman} {et~al.}(2002){Freeman}, {Kashyap}, {Rosner}, \&
  {Lamb}}]{f02}
{Freeman}, P.~E., {Kashyap}, V., {Rosner}, R., \& {Lamb}, D.~Q. 2002, \apjs,
  138, 185

\bibitem[{{Gordon} {et~al.}(2008){Gordon}, {Engelbracht}, {Rieke}, {Misselt},
  {Smith}, \& {Kennicutt}}]{gordon}
{Gordon}, K.~D., {Engelbracht}, C.~W., {Rieke}, G.~H., {Misselt}, K.~A.,
  {Smith}, J., \& {Kennicutt}, Jr., R.~C. 2008, \apj, 682, 336

\bibitem[{{Guhathakurta} \& {Draine}(1989)}]{raja}
{Guhathakurta}, P., \& {Draine}, B.~T. 1989, \apj, 345, 230

\bibitem[{{Harper-Clark} \& {Murray}(2009)}]{hc09}
{Harper-Clark}, E., \& {Murray}, N. 2009, \apj, 693, 1696

\bibitem[{{Hartmann} {et~al.}(2001){Hartmann}, {Ballesteros-Paredes}, \&
  {Bergin}}]{hartmann}
{Hartmann}, L., {Ballesteros-Paredes}, J., \& {Bergin}, E.~A. 2001, \apj, 562,
  852

\bibitem[{{Haynes} {et~al.}(1991){Haynes}, {Klein}, {Wayte}, {Wielebinski},
  {Murray}, {Bajaja}, {Meinert}, {Buczilowski}, {Harnett}, {Hunt}, {Wark}, \&
  {Sciacca}}]{haynes}
{Haynes}, R.~F., {et~al.} 1991, \aap, 252, 475

\bibitem[{{Helou} {et~al.}(2004){Helou}, {Roussel}, {Appleton}, {Frayer},
  {Stolovy}, {Storrie-Lombardi}, {Hurt}, {Lowrance}, {Makovoz}, {Masci},
  {Surace}, {Gordon}, {Alonso-Herrero}, {Engelbracht}, {Misselt}, {Rieke},
  {Rieke}, {Willner}, {Pahre}, {Ashby}, {Fazio}, \& {Smith}}]{helou}
{Helou}, G., {et~al.} 2004, \apjs, 154, 253

\bibitem[{{Hodge} {et~al.}(1994{\natexlab{a}}){Hodge}, {Kennicutt}, \&
  {Strobel}}]{hks94}
{Hodge}, P., {Kennicutt}, R.~C., \& {Strobel}, N. 1994{\natexlab{a}}, \pasp,
  106, 765

\bibitem[{{Hodge} {et~al.}(1989{\natexlab{a}}){Hodge}, {Lee}, \&
  {Kennicutt}}]{hlk2}
{Hodge}, P., {Lee}, M.~G., \& {Kennicutt}, Jr., R.~C. 1989{\natexlab{a}},
  \pasp, 101, 640

\bibitem[{{Hodge} {et~al.}(1989{\natexlab{b}}){Hodge}, {Lee}, \&
  {Kennicutt}}]{hlk89}
---. 1989{\natexlab{b}}, \pasp, 101, 32

\bibitem[{{Hodge} {et~al.}(1994{\natexlab{b}}){Hodge}, {Strobel}, \&
  {Kennicutt}}]{hsk94}
{Hodge}, P., {Strobel}, N.~V., \& {Kennicutt}, R.~C. 1994{\natexlab{b}}, \pasp,
  106, 309

\bibitem[{{Hodge} {et~al.}(1999){Hodge}, {Balsley}, {Wyder}, \&
  {Skelton}}]{h99}
{Hodge}, P.~W., {Balsley}, J., {Wyder}, T.~K., \& {Skelton}, B.~P. 1999, \pasp,
  111, 685

\bibitem[{{Hunter} {et~al.}(1995){Hunter}, {Shaya}, {Holtzman}, {Light},
  {O'Neil}, \& {Lynds}}]{hunter95}
{Hunter}, D.~A., {Shaya}, E.~J., {Holtzman}, J.~A., {Light}, R.~M., {O'Neil},
  Jr., E.~J., \& {Lynds}, R. 1995, \apj, 448, 179

\bibitem[{{Indebetouw} {et~al.}(2009){Indebetouw}, {de Messi{\`e}res},
  {Madden}, {Engelbracht}, {Smith}, {Meixner}, {Brandl}, {Smith}, {Boulanger},
  {Galliano}, {Gordon}, {Hora}, {Sewilo}, {Tielens}, {Werner}, \&
  {Wolfire}}]{remy}
{Indebetouw}, R., {et~al.} 2009, \apj, 694, 84

\bibitem[{{Jijina} \& {Adams}(1996)}]{jijina}
{Jijina}, J., \& {Adams}, F.~C. 1996, \apj, 462, 874

\bibitem[{{Johansson} {et~al.}(1998){Johansson}, {Greve}, {Booth}, {Boulanger},
  {Garay}, {de Graauw}, {Israel}, {Kutner}, {Lequeux}, {Murphy}, {Nyman}, \&
  {Rubio}}]{johansson}
{Johansson}, L.~E.~B., {et~al.} 1998, \aap, 331, 857

\bibitem[{{Kennicutt}(1984)}]{kenn84}
{Kennicutt}, Jr., R.~C. 1984, \apj, 287, 116

\bibitem[{{Kennicutt} \& {Hodge}(1986)}]{kenn86}
{Kennicutt}, Jr., R.~C., \& {Hodge}, P.~W. 1986, \apj, 306, 130

\bibitem[{{Kennicutt} {et~al.}(2008){Kennicutt}, {Lee}, {Funes}, {Sakai}, \&
  {Akiyama}}]{k08}
{Kennicutt}, Jr., R.~C., {Lee}, J.~C., {Funes}, Jos{\'e}~G., S.~J., {Sakai},
  S., \& {Akiyama}, S. 2008, \apjs, 178, 247

\bibitem[{{Knapen} {et~al.}(2003){Knapen}, {de Jong}, {Stedman}, \&
  {Bramich}}]{knap}
{Knapen}, J.~H., {de Jong}, R.~S., {Stedman}, S., \& {Bramich}, D.~M. 2003,
  \mnras, 344, 527

\bibitem[{{Krabbe} \& {Copetti}(2002)}]{krabbe}
{Krabbe}, A.~C., \& {Copetti}, M.~V.~F. 2002, \aap, 387, 295

\bibitem[{{Krumholz} \& {Matzner}(2009)}]{km09}
{Krumholz}, M.~R., \& {Matzner}, C.~D. 2009, \apj, 703, 1352

\bibitem[{{Krumholz} {et~al.}(2006){Krumholz}, {Matzner}, \& {McKee}}]{mrk06}
{Krumholz}, M.~R., {Matzner}, C.~D., \& {McKee}, C.~F. 2006, \apj, 653, 361

\bibitem[{{Krumholz} \& {Tan}(2007)}]{kt07}
{Krumholz}, M.~R., \& {Tan}, J.~C. 2007, \apj, 654, 304

\bibitem[{{Kudritzki} {et~al.}(1999){Kudritzki}, {Puls}, {Lennon}, {Venn},
  {Reetz}, {Najarro}, {McCarthy}, \& {Herrero}}]{kud}
{Kudritzki}, R.~P., {Puls}, J., {Lennon}, D.~J., {Venn}, K.~A., {Reetz}, J.,
  {Najarro}, F., {McCarthy}, J.~K., \& {Herrero}, A. 1999, \aap, 350, 970

\bibitem[{{Kurt} \& {Dufour}(1998)}]{kd98}
{Kurt}, C.~M., \& {Dufour}, R.~J. 1998, in Revista Mexicana de Astronomia y
  Astrofisica, vol. 27, Vol.~7, Revista Mexicana de Astronomia y Astrofisica
  Conference Series, ed. {R.~J.~Dufour \& S.~Torres-Peimbert}, 202--+

\bibitem[{{Lanz} \& {Hubeny}(2003)}]{lanz}
{Lanz}, T., \& {Hubeny}, I. 2003, \apjs, 146, 417

\bibitem[{{Lazendic} {et~al.}(2003){Lazendic}, {Dickel}, \& {Jones}}]{lazendic}
{Lazendic}, J.~S., {Dickel}, J.~R., \& {Jones}, P.~A. 2003, \apj, 596, 287

\bibitem[{{Leitherer}(1997)}]{mini}
{Leitherer}, C. 1997, in Revista Mexicana de Astronomia y Astrofisica, vol. 27,
  Vol.~6, Revista Mexicana de Astronomia y Astrofisica Conference Series, ed.
  {J.~Franco, R.~Terlevich, \& A.~Serrano}, 114--+

\bibitem[{{Leitherer} {et~al.}(1999){Leitherer}, {Schaerer}, {Goldader},
  {Gonz{\'a}lez Delgado}, {Robert}, {Kune}, {de Mello}, {Devost}, \&
  {Heckman}}]{starburst99}
{Leitherer}, C., {et~al.} 1999, \apjs, 123, 3

\bibitem[{{Li} \& {Nakamura}(2006)}]{li}
{Li}, Z., \& {Nakamura}, F. 2006, \apjl, 640, L187

\bibitem[{{Liedahl} {et~al.}(1995){Liedahl}, {Osterheld}, \& {Goldstein}}]{l95}
{Liedahl}, D.~A., {Osterheld}, A.~L., \& {Goldstein}, W.~H. 1995, \apjl, 438,
  L115

\bibitem[{{Malumuth} \& {Heap}(1994)}]{mal94}
{Malumuth}, E.~M., \& {Heap}, S.~R. 1994, \aj, 107, 1054

\bibitem[{{Massey} \& {Hunter}(1998)}]{mh98}
{Massey}, P., \& {Hunter}, D.~A. 1998, \apj, 493, 180

\bibitem[{{Mathis} {et~al.}(1983){Mathis}, {Mezger}, \& {Panagia}}]{mathis}
{Mathis}, J.~S., {Mezger}, P.~G., \& {Panagia}, N. 1983, \aap, 128, 212

\bibitem[{{Matzner}(2002)}]{m02}
{Matzner}, C.~D. 2002, \apj, 566, 302

\bibitem[{{Matzner}(2007)}]{matzner07}
---. 2007, \apj, 659, 1394

\bibitem[{{McKee} \& {Williams}(1997)}]{mw97}
{McKee}, C.~F., \& {Williams}, J.~P. 1997, \apj, 476, 144

\bibitem[{{Meixner} {et~al.}(2006){Meixner}, {Gordon}, {Indebetouw}, {Hora},
  {Whitney}, {Blum}, {Reach}, {Bernard}, {Meade}, {Babler}, {Engelbracht},
  {For}, {Misselt}, {Vijh}, {Leitherer}, {Cohen}, {Churchwell}, {Boulanger},
  {Frogel}, {Fukui}, {Gallagher}, {Gorjian}, {Harris}, {Kelly}, {Kawamura},
  {Kim}, {Latter}, {Madden}, {Markwick-Kemper}, {Mizuno}, {Mizuno}, {Mould},
  {Nota}, {Oey}, {Olsen}, {Onishi}, {Paladini}, {Panagia}, {Perez-Gonzalez},
  {Shibai}, {Sato}, {Smith}, {Staveley-Smith}, {Tielens}, {Ueta}, {van Dyk},
  {Volk}, {Werner}, \& {Zaritsky}}]{meixner}
{Meixner}, M., {et~al.} 2006, \aj, 132, 2268

\bibitem[{{Meurer} {et~al.}(1997){Meurer}, {Heckman}, {Lehnert}, {Leitherer},
  \& {Lowenthal}}]{meurer}
{Meurer}, G.~R., {Heckman}, T.~M., {Lehnert}, M.~D., {Leitherer}, C., \&
  {Lowenthal}, J. 1997, \aj, 114, 54

\bibitem[{{Mewe} {et~al.}(1985){Mewe}, {Gronenschild}, \& {van den Oord}}]{m85}
{Mewe}, R., {Gronenschild}, E.~H.~B.~M., \& {van den Oord}, G.~H.~J. 1985,
  \aaps, 62, 197

\bibitem[{{Mewe} {et~al.}(1986){Mewe}, {Lemen}, \& {van den Oord}}]{m86}
{Mewe}, R., {Lemen}, J.~R., \& {van den Oord}, G.~H.~J. 1986, \aaps, 65, 511

\bibitem[{{Motte} {et~al.}(1998){Motte}, {Andre}, \& {Neri}}]{motte}
{Motte}, F., {Andre}, P., \& {Neri}, R. 1998, \aap, 336, 150

\bibitem[{{Murray} {et~al.}(2010{\natexlab{a}}){Murray}, {M{\'e}nard}, \&
  {Thompson}}]{wind}
{Murray}, N., {M{\'e}nard}, B., \& {Thompson}, T.~A. 2010{\natexlab{a}}, ArXiv
  e-prints

\bibitem[{{Murray} {et~al.}(2010{\natexlab{b}}){Murray}, {Quataert}, \&
  {Thompson}}]{murray}
{Murray}, N., {Quataert}, E., \& {Thompson}, T.~A. 2010{\natexlab{b}}, \apj,
  709, 191

\bibitem[{{Murray} \& {Rahman}(2010)}]{mr10}
{Murray}, N., \& {Rahman}, M. 2010, \apj, 709, 424

\bibitem[{{Nakamura} \& {Li}(2008)}]{naka}
{Nakamura}, F., \& {Li}, Z. 2008, \apj, 687, 354

\bibitem[{{Onishi} {et~al.}(2002){Onishi}, {Mizuno}, {Kawamura}, {Tachihara},
  \& {Fukui}}]{onishi}
{Onishi}, T., {Mizuno}, A., {Kawamura}, A., {Tachihara}, K., \& {Fukui}, Y.
  2002, \apj, 575, 950

\bibitem[{{Parker}(1993)}]{parker1}
{Parker}, J.~W. 1993, \aj, 106, 560

\bibitem[{{Parker} \& {Garmany}(1993)}]{parker2}
{Parker}, J.~W., \& {Garmany}, C.~D. 1993, \aj, 106, 1471

\bibitem[{{Peeters} {et~al.}(2004){Peeters}, {Spoon}, \& {Tielens}}]{peeters}
{Peeters}, E., {Spoon}, H.~W.~W., \& {Tielens}, A.~G.~G.~M. 2004, \apj, 613,
  986

\bibitem[{{Quillen} {et~al.}(2005){Quillen}, {Thorndike}, {Cunningham},
  {Frank}, {Gutermuth}, {Blackman}, {Pipher}, \& {Ridge}}]{quill}
{Quillen}, A.~C., {Thorndike}, S.~L., {Cunningham}, A., {Frank}, A.,
  {Gutermuth}, R.~A., {Blackman}, E.~G., {Pipher}, J.~L., \& {Ridge}, N. 2005,
  \apj, 632, 941

\bibitem[{{Repolust} {et~al.}(2004){Repolust}, {Puls}, \& {Herrero}}]{rep}
{Repolust}, T., {Puls}, J., \& {Herrero}, A. 2004, \aap, 415, 349

\bibitem[{{Rieke} {et~al.}(2004){Rieke}, {Young}, {Engelbracht}, {Kelly},
  {Low}, {Haller}, {Beeman}, {Gordon}, {Stansberry}, {Misselt}, {Cadien},
  {Morrison}, {Rivlis}, {Latter}, {Noriega-Crespo}, {Padgett}, {Stapelfeldt},
  {Hines}, {Egami}, {Muzerolle}, {Alonso-Herrero}, {Blaylock}, {Dole}, {Hinz},
  {Le Floc'h}, {Papovich}, {P{\'e}rez-Gonz{\'a}lez}, {Smith}, {Su}, {Bennett},
  {Frayer}, {Henderson}, {Lu}, {Masci}, {Pesenson}, {Rebull}, {Rho}, {Keene},
  {Stolovy}, {Wachter}, {Wheaton}, {Werner}, \& {Richards}}]{rieke}
{Rieke}, G.~H., {et~al.} 2004, \apjs, 154, 25

\bibitem[{{Selman} \& {Melnick}(2005)}]{selman05}
{Selman}, F.~J., \& {Melnick}, J. 2005, \aap, 443, 851

\bibitem[{{Shapley} {et~al.}(2003){Shapley}, {Steidel}, {Pettini}, \&
  {Adelberger}}]{shapley}
{Shapley}, A.~E., {Steidel}, C.~C., {Pettini}, M., \& {Adelberger}, K.~L. 2003,
  \apj, 588, 65

\bibitem[{{Smith} \& {MCELS Team}(1998)}]{smith}
{Smith}, R.~C., \& {MCELS Team}. 1998, Publications Astronomical Society of
  Australia, 15, 163

\bibitem[{{Strobel} {et~al.}(1990){Strobel}, {Hodge}, \& {Kennicutt}}]{shk90}
{Strobel}, N.~V., {Hodge}, P., \& {Kennicutt}, Jr., R.~C. 1990, \pasp, 102,
  1241

\bibitem[{{Strobel} {et~al.}(1991){Strobel}, {Hodge}, \& {Kennicutt}}]{shk91}
---. 1991, \apj, 383, 148

\bibitem[{{Tan} {et~al.}(2006){Tan}, {Krumholz}, \& {McKee}}]{tan06}
{Tan}, J.~C., {Krumholz}, M.~R., \& {McKee}, C.~F. 2006, \apjl, 641, L121

\bibitem[{{Testi} \& {Sargent}(1998)}]{ts98}
{Testi}, L., \& {Sargent}, A.~I. 1998, \apjl, 508, L91

\bibitem[{{Thompson} {et~al.}(2005){Thompson}, {Quataert}, \&
  {Murray}}]{thompson}
{Thompson}, T.~A., {Quataert}, E., \& {Murray}, N. 2005, \apj, 630, 167

\bibitem[{{Townsley} {et~al.}(2006{\natexlab{a}}){Townsley}, {Broos},
  {Feigelson}, {Brandl}, {Chu}, {Garmire}, \& {Pavlov}}]{townsley1}
{Townsley}, L.~K., {Broos}, P.~S., {Feigelson}, E.~D., {Brandl}, B.~R., {Chu},
  Y., {Garmire}, G.~P., \& {Pavlov}, G.~G. 2006{\natexlab{a}}, \aj, 131, 2140

\bibitem[{{Townsley} {et~al.}(2006{\natexlab{b}}){Townsley}, {Broos},
  {Feigelson}, {Garmire}, \& {Getman}}]{townsley2}
{Townsley}, L.~K., {Broos}, P.~S., {Feigelson}, E.~D., {Garmire}, G.~P., \&
  {Getman}, K.~V. 2006{\natexlab{b}}, \aj, 131, 2164

\bibitem[{{Walborn}(1991)}]{wal91}
{Walborn}, N.~R. 1991, in IAU Symposium, Vol. 148, The Magellanic Clouds, ed.
  {R.~Haynes \& D.~Milne}, 145--+

\bibitem[{{Walborn} \& {Blades}(1997)}]{wb97}
{Walborn}, N.~R., \& {Blades}, J.~C. 1997, \apjs, 112, 457

\bibitem[{{Wang} {et~al.}(2010){Wang}, {Li}, {Abel}, \& {Nakamura}}]{wang}
{Wang}, P., {Li}, Z., {Abel}, T., \& {Nakamura}, F. 2010, \apj, 709, 27

\bibitem[{{Wang}(1999)}]{wang99}
{Wang}, Q.~D. 1999, \apjl, 510, L139

\bibitem[{{Weaver} {et~al.}(1977){Weaver}, {McCray}, {Castor}, {Shapiro}, \&
  {Moore}}]{weaver77}
{Weaver}, R., {McCray}, R., {Castor}, J., {Shapiro}, P., \& {Moore}, R. 1977,
  \apj, 218, 377

\bibitem[{{Weingartner} \& {Draine}(2001)}]{wd01}
{Weingartner}, J.~C., \& {Draine}, B.~T. 2001, \apjs, 134, 263

\bibitem[{{Whitworth}(1979)}]{whit}
{Whitworth}, A. 1979, \mnras, 186, 59

\bibitem[{{Williams} \& {McKee}(1997)}]{wk97}
{Williams}, J.~P., \& {McKee}, C.~F. 1997, \apj, 476, 166

\bibitem[{{Wyder} {et~al.}(1997){Wyder}, {Hodge}, \& {Skelton}}]{wyder}
{Wyder}, T.~K., {Hodge}, P.~W., \& {Skelton}, B.~P. 1997, \pasp, 109, 927

\bibitem[{{Yorke} {et~al.}(1989){Yorke}, {Tenorio-Tagle}, {Bodenheimer}, \&
  {Rozyczka}}]{yorke}
{Yorke}, H.~W., {Tenorio-Tagle}, G., {Bodenheimer}, P., \& {Rozyczka}, M. 1989,
  \aap, 216, 207

\bibitem[{{Zuckerman} \& {Evans}(1974)}]{zuckerman}
{Zuckerman}, B., \& {Evans}, II, N.~J. 1974, \apjl, 192, L149

\end{thebibliography}

\end{document}